\documentclass[twocolumn]{aastex631}

\usepackage[version=4]{mhchem}

\begin{document}

\title{Carbon Isotope Fractionation of Complex Organic Molecules in Star-Forming Cores}

\correspondingauthor{Ryota Ichimura}
\email{ryotaichimura.astrolife@gmail.com}

\author[0009-0002-3208-3296]{RYOTA ICHIMURA}
\affiliation{Division of Science, National Astronomical Observatory of Japan, 2-21-1 Osawa, Mitaka, Tokyo 181-8588, Japan}
\affiliation{Department of Astronomical Science, The Graduate University for Advanced Studies, SOKENDAI, 2-21-1 Osawa, Mitaka, Tokyo 181-8588, Japan}

\author[0000-0002-7058-7682]{HIDEKO NOMURA}
\affiliation{Division of Science, National Astronomical Observatory of Japan, 2-21-1 Osawa, Mitaka, Tokyo 181-8588, Japan}
\affiliation{Department of Astronomical Science, The Graduate University for Advanced Studies, SOKENDAI, 2-21-1 Osawa, Mitaka, Tokyo 181-8588, Japan}

\author[0000-0002-2026-8157]{KENJI FURUYA}
\affiliation{Division of Science, National Astronomical Observatory of Japan, 2-21-1 Osawa, Mitaka, Tokyo 181-8588, Japan}
\affiliation{Department of Astronomy, Graduate School of Science, University of Tokyo, Tokyo 113-0033, Japan}

\begin{abstract}

Recent high-resolution and sensitivity ALMA observations have unveiled the carbon isotope ratios ($^{12}$C/$^{13}$C) of Complex Organic Molecules (COMs) in a low-mass protostellar source.
To understand the $^{12}$C/$^{13}$C ratios of COMs, we investigated the carbon isotope fractionation of COMs from prestellar cores to protostellar cores with a gas-grain chemical network model. We confirmed that the $^{12}$C/$^{13}$C ratios of small molecules are bimodal in the prestellar phase: CO and species formed from CO (e.g., CH$_{3}$OH) are slightly enriched in $^{13}$C compared to the local ISM (by $\sim$ 10 $\%$), while those from C and C$^{+}$ are depleted in $^{13}$C owing to isotope exchange reactions. 
COMs are mainly formed on the grain surface and in the hot gas ($>$ 100 K) in the protostellar phase. The $^{12}$C/$^{13}$C ratios of COMs depend on which molecules the COMs are formed from.
In our base model, some COMs in the hot gas are depleted in $^{13}$C compared to the observations. Thus,
We additionally incorporate reactions between gaseous atomic C and H$_{2}$O ice or CO ice on the grain surface to form H$_2$CO ice or \ce{C2O} ice, as suggested by recent laboratory studies. The direct C-atom addition reactions open pathways to form \ce{^13C}-enriched COMs from atomic C and CO ice. We find that these direct C-atom addition reactions mitigate $^{13}$C-depletion of COMs, and the model with the direct C-atom addition reactions better reproduces the observations than our base model. We also discuss the impact of the cosmic ray ionization rate on the $^{12}$C/$^{13}$C ratio of COMs.

\textbf{}

\end{abstract}

\section{Introduction}
In the central ($<$ 100 AU) and hot ($>$ 100 K) regions in low-mass protostellar cores, rotational transition lines from various species have been observed. 
Some species are referred to as complex organic molecules (COMs), defined as organic molecules with six or more atoms\citep{2009H&D}. 

According to astrochemical models, the formation of COMs requires a sequence of grain surface processes during star formation. This includes freeze-out of gaseous species onto the grain surface, and grain surface chemistry at low temperature ($\sim$ 10 K) to produce simple icy molecules such as CH$_{4}$ and CH$_{3}$OH. 
After protostar formation, the gas and dust temperatures increase (above $\sim$ 25 K), radicals form and COMs are produced via radical-radical reactions on the grain surface via diffusion(e.g., \citealp{2006Garrod,2009H&D}).
Additionally, recent laboratory experiments suggest that the atomic C insertion or addition reaction on the grain surface may be important for the production of the COMs and increasing their complexity \citep{2023Tsuge, 2024ApJF}. Also, nondiffusive grain surface chemistry may be important for the formation of COMs under low-temperature conditions \citep{2018ApJShingledecker,2020Jin,2022Garrod}. 
However, it is still challenging to reveal the formation pathways of COMs.

The measurement of isotope ratios and their fractionation is useful to investigate the local chemical process such as the formation pathways of COMs in star-forming regions.
The carbon isotope ratio of different COMs has been measured with high-resolution ALMA observations towards the Class 0 low-mass protostellar binary system IRAS 16293-2422. The values were found to be comparable with the local ISM value ($^{12}$C/$^{13}$C = 68; \citealp{2005Milam}) or to show lower $^{12}$C/$^{13}$C ratios towards source B \citep{2016Jorgensen,2018Jorgensen}, while are consistent within the error with the local ISM value for source A \citep{2020Manigand}. In the Class I low-mass young outbursting star V883 Ori, COMs are enriched in $^{13}$C ($^{12}$C/$^{13}$C $\sim$ 20-30) \citep{2024Yamato}.

Carbon isotope fractionation in the ISM is thought to occur via exothermic isotope exchange reactions, isotope selective photodissociation, and desorption from grain surface with the different binding energies of isotopologues \citep{2023Nomura}. 
Especially, several studies of astrochemical models focus on exothermic isotope exchange reactions to reproduce the observed carbon isotope ratios of fractionated simple molecules (e.g. \ce{C2H}/\ce{^13CCH} $>$ 250 and \ce{HC3N}/\ce{H^13CCCN} = 79) in dense interstellar clouds \citep{langer1984,furuya2011,2015Roueff,colzi2020,loison2020,2023Sipila}. Isotope exchange reactions can fractionate carbon-bearing species because of the zero-point vibrational energy difference between $^{13}$C isotopologues and $^{12}$C isotopologues. One of the most important carbon isotope exchange reactions is 
\begin{equation}
    \label{eq:1}
    \rm{^{13}C^{+}+{}^{12}CO\rightleftharpoons{}^{12}C^{+}+{}^{13}CO + \Delta E, }
\end{equation}
where the vibrational energy difference is $\Delta$ E = 35 K \citep{watson1976}. \citet{langer1984} have introduced this exchange reaction into their gas-phase astrochemical model. They concluded that at a low temperature ($\sim$20 K), the forward reaction of the reaction in Eq.(\ref{eq:1}) is efficient compared to the backward reaction, and the small carbon-bearing species are divided into two groups; $^{13}$C-rich species formed from CO (e.g., CO$_{2}$ and CH$_{3}$OH) and $^{13}$C-poor species formed from C$^{+}$ (e.g., CH$_{4}$). In addition to the reaction in Eq.(\ref{eq:1}), other carbon isotope exchange reactions have been suggested based on quantum chemical calculations such as $^{13}$C + C$_3$ $\rightleftharpoons$ $^{12}$C + $^{13}$CC$_2$ + 28 K and lead to carbon isotope fractionation \citep{2015Roueff,colzi2020,loison2020}.

Isotope selective photodissociation and the difference in binding energies among different isotopologues also cause carbon isotope fractionation. For example, self-shielding of CO against UV photodissociation can cause the fractionation \citep{1988Ewine,2009Visser}. This effect is however only pronounced at visual extinction of Av $\sim$1 mag. Moreover, $^{12}$CO have a slightly lower binding energy than $^{13}$CO because it is lighter \citep{2015Smith,2021Smith}. Consequently, carbon-bearing species in the solid phase may be enriched in $^{13}$C, while those in the gas phase may be depleted in $^{13}$C.
\citet{2016Jorgensen,2018Jorgensen} suggested that the observed carbon isotope fractionation of COMs in IRAS16293-2422B might be affected by isotope selective photodissociation of CO and/or difference in binding energies between $^{12}$CO and $^{13}$CO.

\citet{furuya2011,colzi2020,loison2020,2023Sipila} studied carbon isotope fractionation of some simple molecules such as \ce{C2H} and \ce{HCN} and \ce{HC3N} using the astrochemical model in prestellar cores with 10K. Meanwhile, \citet{garrod2013, aikawa2008} studied the formation of COMs using astrochemical models in low- and high-mass star-forming cores with temperatures up to around 200K.

In this paper, we report gas-grain chemical reaction network calculations, systematically investigating the carbon isotope fractionations of COMs. This calculation includes their formation before star formation and their sublimation into the gas phase following star formation. In section \ref{sec: Model}, we describe a physical model of a star-forming core and the gas-grain chemical model. In Section \ref{sec: The base Model}, we present the results of time variation of molecular abundances and $^{12}$C/$^{13}$C ratios in the base model. The effects of the atomic C insertion or addition reactions on the grain surface and difference in binding energy between $^{12}$CO and $^{13}$CO and the initial condition of carbon and $^{12}$C/$^{13}$C ratio of C$^{+}$ are also studied. In Section \ref{sec: Discussion}, we compare our calculation results with the observations in the IRAS 16293-2422B source. We also discuss the effect of cosmic ray (CR) ionization rates on $^{12}$C/$^{13}$C ratios of COMs. Our findings are summarized in Section \ref{sec: sum}.

\section{Model} \label{sec: Model}
\subsection{Physical Model}

\begin{figure}[ht!]
\plotone{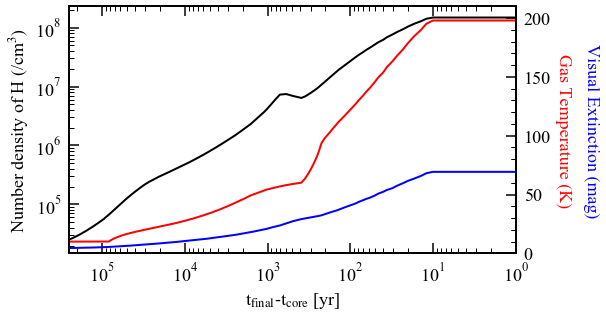}
\caption{Temporal variation of the number density of hydrogen nuclei (black line), gas temperature (red line), and visual extinction (blue line) of a fluid parcel along a streamline in a gravitationally collapsing core.
\label{fig: physicalmodel}}
\end{figure}

We adopt one-dimensional radiation hydrodynamics simulations by \citet{m&i2000} as a physical model. This model traces the evolution of a dense, starless core to a protostellar core.
A fluid parcel is traced in the hydrodynamics simulation, and we calculate the evolution of molecular abundances with a chemical reaction network along the fluid parcel as done in \citet{aikawa2008,aikawa2020}.
The model is divided into two phases: the static phase and the collapse phase. In the static phase, the prestellar core is in hydrostatic equilibrium assuming that turbulence prevents the contraction of the core. In the calculations, we adopt the gas and dust temperature of 10 K, the number density of hydrogen nuclei of 2.28$\times$10$^4$ cm$^{-3}$, and the visual extinction of 4.51 mag, which correspond to the value of the fluid parcel at 1$\times$10$^4$ au from the core center \citep{m&i2000,aikawa2008}.
In the subsequent collapse phase, a protostar is formed due to gravitational collapse and further grows through envelope accretion. Following the first collapse and the formation of
an adiabatic core, the protostar is born at 2.5$\times$10$^5$ yr when the second collapse stops \citep{m&i2000,aikawa2008}. This model further proceeds to track the physical evolution for 9.3$\times$10$^4$ yr until the fluid parcel reaches 30.6 au from the core center.
Figure \ref{fig: physicalmodel} shows the temporal variation of the number density of hydrogen nuclei, gas temperature, and visual extinction in a fluid parcel. We define the moment of the protostar formation as t$\mathrm{_{core}}$ = 0 and the end of calculation as t$\mathrm{_{core}}$ = t$\mathrm{_{final}}$ = 9.3$\times$10$^4$ yr. To increase the visibility of the evolution around the final stage, we adjust the horizontal axis to the logarithmic scale of t$\mathrm{_{final}}$ - t$\mathrm{_{core}}$ \citep{aikawa2008}.
At t$\mathrm{_{core}}$ = t$\mathrm{_{final}}$ in the collapse phase, which is the final time step with both gas and dust temperatures reaching $\sim$ 198 K, the total H$_2$ density reaching $\sim$ 1.49$\times$10$^8$ cm$^{-3}$, and visual extinction $\sim$ 70 mag. 
The infalling timescale inside 1000 au is shorter than the lifetime of the protostar in this model so that we can use the time evolution of abundances in a single fluid parcel to represent the true radial distribution \citep{aikawa2020}. For the sake of simplicity, we set a minimum temperature of 10 K throughout our simulations.

\subsection{Chemical Model}  \label{sec: chemical model}
\subsubsection{The base model}
We utilize the gas-ice astrochemical code based on the rate equation approach (Rokko code; \citealp{2015Furuya}) and incorporate two hydrogenation reactions on grain surface; CH$_2$CO ice + H ice $\to$ CH$_3$CO ice and CH$_3$CO ice + H ice $\to$ CH$_3$CHO \citep{2015Ruaud}.
We also expand to include mono-$^{13}$C species and carbon isotope exchange reactions (see Table \ref{table: exchange}). 
The chemistry is described by a three-phase model \citep{furuya2016}. This model makes a distinction between the surface of the ice mantle and the rest of the ice mantle \citep{h&h1993}. Like in \citet{h&h1993}, we assume that the ice mantle phase remains chemically inert, while the ice surface phase considers chemical reactions. 
Species in our chemical network consist of carbon skeletons up to three carbons to reduce computational time.
We consider the species with a different position of $^{13}$C as the same species (e.g., $^{12}$C$^{13}$CH and $^{13}$C$^{12}$CH are the same species) for simplicity.
We have omitted the multiple fractionations, such as two or more $^{13}$C for simplicity. Isotope exchange reactions are taken from \citet{2015Roueff} and \citet{loison2020}, and they are considered in the gas phase. $^{13}$C-bearing species, however, are taken into account in ice surface chemistry through the adsorption of $^{13}$C-bearing species onto the grain surface and the subsequent grain surface reactions, such as hydrogenation of species.
We neglect $^{13}$C fractionation via the isotope selective photodissociation in the following sections because we only consider the dense region.
Our reaction network consists of 733 gas and grain species and 20360 gas phase and grain surface reactions. 

We assume the Langmuir-Hinshelwood mechanism to describe two-body reactions on grain surfaces. In this mechanism, species on the grain surface diffuse by thermal hopping and react with each other when they meet. The set of adsorption energies is adopted from \citet{2015Furuya}. The barrier for thermal diffusion of atoms and molecules, excluding hydrogen (H), is set to be 40 $\%$ of the adsorption energy ($E_{des}$). The barrier is set at 30 $\%$ of $E_{des}$ for hydrogen. We assume that the sticking probability of colliding gaseous species except for hydrogen onto the grain is unity.
For hydrogen, the sticking probability is calculated based on \citet{1979Hollenbach}. 
We adopt the low-metal elemental abundances taken from \citet{aikawa2001} through our calculations. The species are assumed to be initially atoms or atomic ions, except for hydrogen, which is in H$_2$.
The dust grain is spherical with a 0.1 {\textmu}m radius with the material density of 2.5 g/cm$^3$. The dust-to-gas ratio is set to 0.01. The CR ionization rate of H$_2$ is set to be 1.3$\times$10$^{-17}$ s$^{-1}$ \citep{1998T&H}. In this paper, we define "abundance" as the fractional abundance of species to hydrogen nuclei. 
$^{12}$C/$^{13}$C ratios are measured by the statistical factor corresponding to the two or three indistinguishable carbon atoms.
For example, \ce{C2H} has two kinds of carbon isotopologues of $^{13}$C$^{12}$CH and $^{12}$C$^{13}$CH, but they are not distinguished in our chemical reaction network. Therefore, the $^{12}$C/$^{13}$C ratio of \ce{C2H} is calculated as ‘n(\ce{C2H})/n($^{13}$C-containung \ce{C2H})’ multiplied by 2 in this work.

\subsubsection{The direct C-atom addition reactions}
Besides the Langmuir-Hinshelwood process, complex molecules could be formed via nondiffusive grain surface reactions at low temperature \citep{2020Jin, 2022Garrod}.
Given atomic carbon's high binding energy of 10,000 K, its diffusion on grain surfaces at the low temperatures of the ISM is challenging. However, recent laboratory experiments and quantum chemical calculations reveal that atomic C reacts with icy light species (e.g., \ce{H2O}, and \ce{CO}) to form \ce{H2CO} and \ce{C2O} ice on the grain surface at 10K, respectively \citep{2021Molp,2022Fedoseev,2023Ferrero}.
Under ISM condition, \ce{H2O} and \ce{CO}, with binding energies of 5,600K and 833K respectively, show slow diffusion on the grain surface. Therefore, we characterize the low-temperature reactions involving atomic C with \ce{H2O} and \ce{CO} chemistries as Eley-Rideal (ER) reactions, which are the direct collision of a gaseous species with adsorbed species on a grain surface.
We incorporate the subsequent two ER reactions involving atomic C (hereafter direct C-atom addition reaction) into the model presented in Section \ref{sec: insertion}. 

\begin{equation}
    \label{eq:2}
    \rm{C + H_2O\,ice \rightarrow H_2CO\,ice.}
\end{equation}

\begin{equation}
    \label{eq:3}
    \rm{C + CO\,ice \rightarrow C_2O\,ice.}
\end{equation}

These direct C-atom addition reactions are assumed to occur when atomic C is adsorbed from the gas phase onto its reaction partner on the grain surface. Therefore, the reaction coefficient (s$^{-1}$) of the direct C-atom addition reaction ($k_{Cadd}$) is given by 
\begin{equation}
    \label{eq:4}
    k_{Cadd} = A \times k_{acc} \times \frac{n_s(X)}{n_s(total)},
\end{equation}

\begin{equation}
    \label{eq:5}
    k_{acc} = n_{grain}S\sigma<v>,
\end{equation}
$k_{acc}$ is the accretion rate coefficient of atomic C onto a grain surface. $n_{grain}$ is the number density of dust grains, \textit{S} is the sticking probability, $\sigma$ is the geometrical cross-section of a dust grain, \textit{$\langle$ v $\rangle$} is the thermal velocity of atomic C, and $n_s(X)$ is the number density of species \textit{X} on a grain surface. For Eq.(\ref{eq:4}),  $n_s(total)$ is the number density of total species present on a grain surface, and \textit{A} is a branching ratio. According to quantum chemical calculations, for the reaction in Eq.(\ref{eq:2}), approximately 30$\%$ of adsorbed atomic C on \ce{H2O} converts to \ce{H2CO} via a barrierless pathway, while the remaining atomic C is present as it is \citep{2021Molp,2023Tsuge}, so the branching ratio of the reaction in Eq.(\ref{eq:2}) is set to be 0.3 (\textit{A} = 0.3).
For the reaction in Eq.(\ref{eq:3}), all adsorbed atomic C on CO converts to \ce{C2O} in a barrierless way \citep{2023Ferrero}, so the branching ratio is set to be 1 (\textit{A} = 1.0).
Some gaseous atomic C take part in these direct C-atom addition reactions instead of the adsorption, so the adsorption rate coefficient of atomic C is adjusted by subtracting the rate coefficient of the direct C-atom addition reaction as 
\begin{equation}
    k_{acc, adj} = k_{acc} - k_{Cadd},
\end{equation}
where $k_{acc, adj}$ is the adjusted accretion rate coefficient.
The impact of direct C-atom addition reactions on the \ce{^12C}/\ce{^13C} ratios of COMs is discussed in Sect. \ref{sec: insertion}.

\begin{figure*}[ht!]
\hspace*{-1.5cm} 
    \begin{tabular}{ccc}
      \begin{minipage}[t]{0.3\hsize}
        \centering
        \includegraphics[keepaspectratio, scale=0.31]{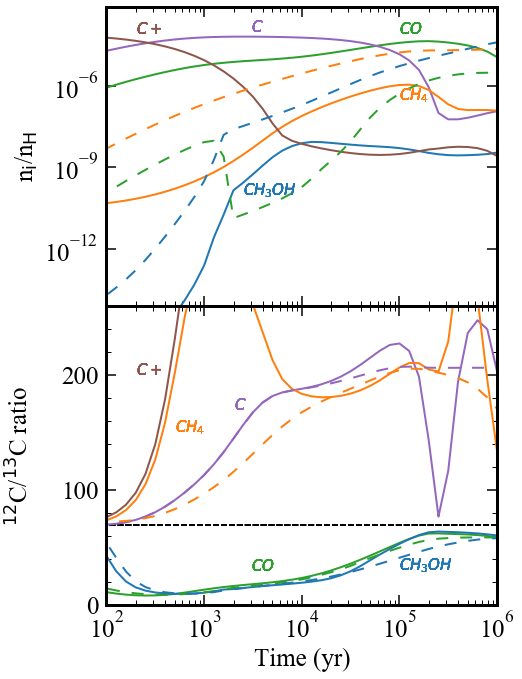}
        \label{composite}
      \end{minipage} &
      \begin{minipage}[t]{0.3\hsize}
        \centering
        \includegraphics[keepaspectratio, scale=0.31]{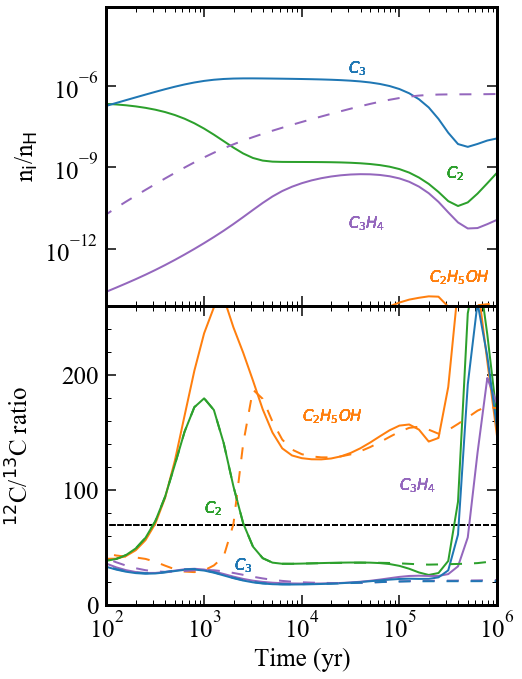}
      \end{minipage} &
      \begin{minipage}[t]{0.3\hsize}
        \centering
        \includegraphics[keepaspectratio, scale=0.31]{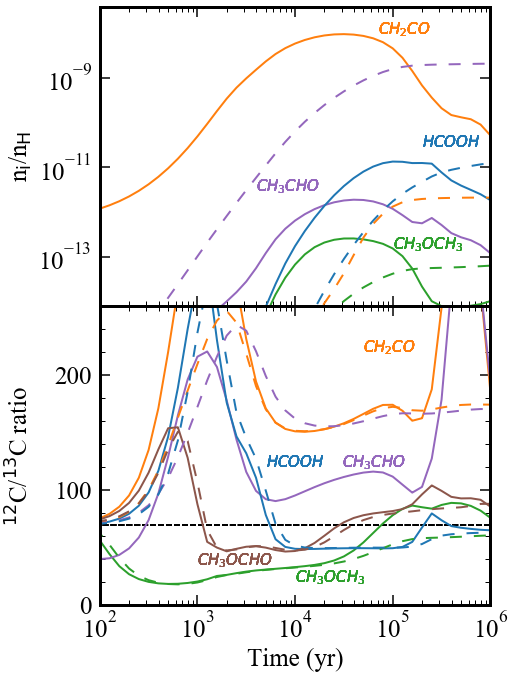}
      \end{minipage}
    \end{tabular}
     \caption{Temporal variation of the molecular abundances and $^{12}$C/$^{13}$C ratios for gaseous species (solid lines) and icy species (dashed lines) during the static phase in the base model. The horizontal black dashed line represents the average $^{12}$C/$^{13}$C ratio of local ISM.}
     \label{fig: static_chemistry}
\end{figure*}

\begin{figure*}[ht!]
\hspace*{-1.4cm} 
    \begin{tabular}{ccc}   
      \begin{minipage}[t]{0.3\hsize}
        \centering
        \includegraphics[keepaspectratio, scale=0.31]{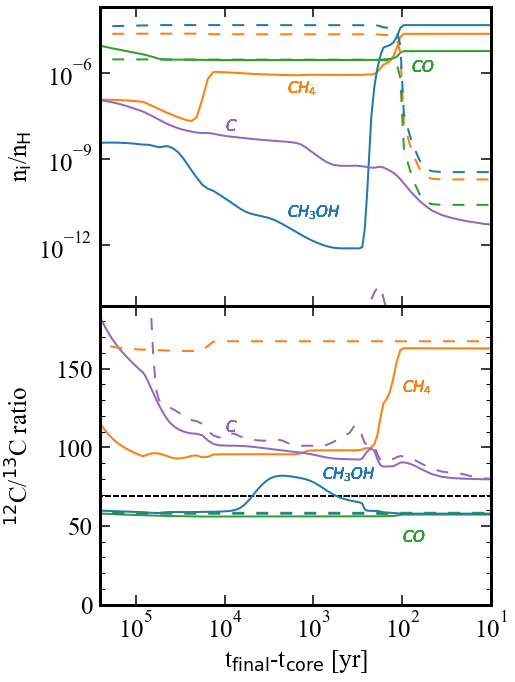}
      \end{minipage} &
      \begin{minipage}[t]{0.3\hsize}
        \centering
        \includegraphics[keepaspectratio, scale=0.31]{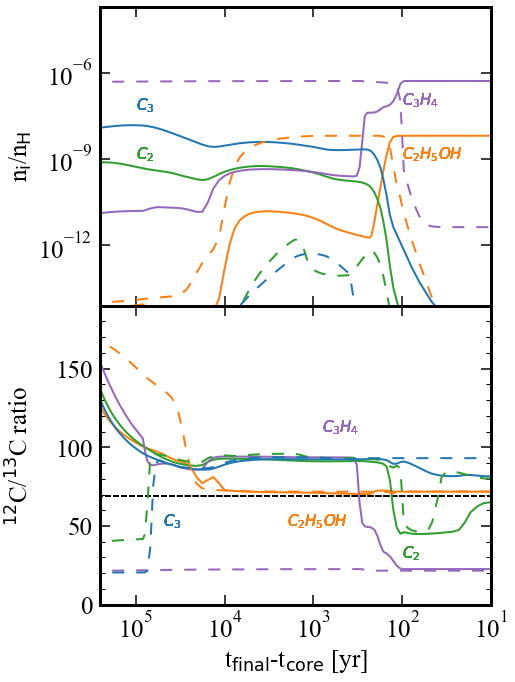}
      \end{minipage} &
      \begin{minipage}[t]{0.3\hsize}
        \centering
        \includegraphics[keepaspectratio, scale=0.31]{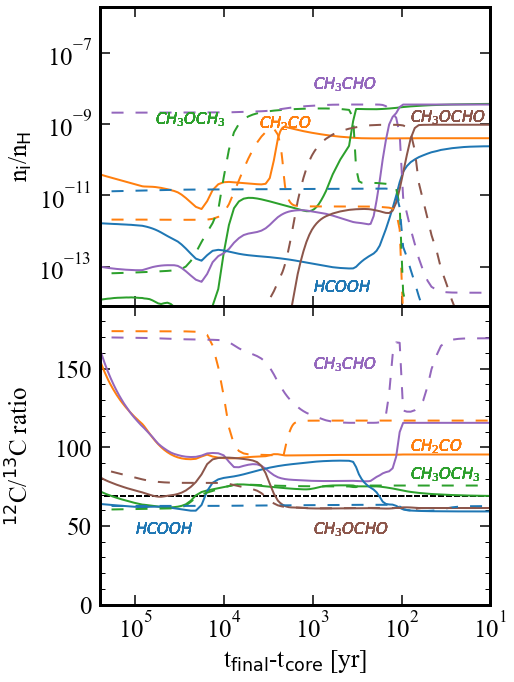}
      \end{minipage} 
    \end{tabular}
     \caption{Same as Figure \ref{fig: static_chemistry} but during the collapse phase.}
     \label{fig: collapse_chemistry}
\end{figure*}

\section{Results} \label{sec: The base Model}
\subsection{The Base Model}

\subsubsection{The Static Phase}
Figure \ref{fig: static_chemistry} shows the temporal variation of abundances and $^{12}$C/$^{13}$C ratios of some selected species in the static phase with a fixed temperature (10 K), density (2.28$\times$10$^{4}$ cm$^{-3}$) and CR ionization rate (1.3$\times$10$^{-17}$ s$^{-1}$). 
With the exception of considering different positions for $^{13}$C or differences in the number of isotope exchange reactions the models in \citet{colzi2020} and \citet{loison2020} are similar to our model in the static phase. Nitrogen-bearing species related to isotope exchange reactions are discussed in Appendix \ref{sec: Nspecies}.
Dominant carriers of carbon in the gas phase change as time goes on. Initially, C$^{+}$ is dominant. C$^{+}$ is gradually converted to atomic C and then to CO at $\>$ $\sim$ 10$^{5}$ yr. 
These small gas-phase species freeze out onto the grain surface and contribute to surface reactions. Adsorbed atomic C is converted to CH$_{4}$ ice via a sequence of hydrogenation reactions on the grain surfaces before 10$^{5}$ yr, while adsorbed CO is converted to CH$_{3}$OH ice.
During the static phase, $^{13}$C fractionation occurs via isotope exchange reactions \citep{furuya2011,colzi2020,loison2020}.
For example, considering $^{13}$C$^{+}$ + $^{12}$CO $\rightleftharpoons$ $^{12}$C$^{+}$ + $^{13}$CO + 35 K  (Eq.(\ref{eq:1})) or $^{13}$C + C$_3$ $\rightleftharpoons$ $^{12}$C + $^{13}$CC$_2$ + 28 K, these isotope exchange reactions lead to the depletion of $^{13}$C in atomic C and C$^{+}$, while CO and C$_3$ become enriched in $^{13}$C until 1 $\times$ 10$^{5}$ yr. 
Around 2 $\times$ 10$^{5}$ yr, CO becomes the dominant carrier of gas-phase carbon, so the $^{12}$C/$^{13}$C ratio of CO gets closer to the local ISM value \citep{furuya2011}. 
The abundance of gaseous atomic C decreases with time due to the conversion to CO and CH$_4$ ice. Around 2 $\times$ 10$^{5}$ yr, gaseous atomic C is produced from C$_3$ and C$_2$ via photodissociation or reaction with atomic O. Therefore, the $^{12}$C/$^{13}$C ratio of atomic C temporally approaches to the local ISM value due to the low $^{12}$C/$^{13}$C ratio of C$_3$ and C$_2$.
After that, C$^{+}$, which is produced through the destruction of CO by \ce{He+}, is converted to gaseous atomic C or C$_3$, so these species become depleted in $^{13}$C.
Around 10$^{3}$ yr, some species such as \ce{C2} are depleted in $^{13}$C as they are formed from C$^{+}$ or atomic C, and after that they are enriched in $^{13}$C since their major reactants change into $^{13}$C-enriched C$_3$ or CO.

Icy molecules (e.g. \ce{CH4} and \ce{CH3OH}) are formed via adsorption of these fractionated small species (e.g. CO or atomic C) and surface reactions, and thus the time variation of $^{12}$C/$^{13}$C ratios of the icy molecules follow those of gas-phase species. For example, before 10$^{5}$ yr, icy molecules such as CH$_{4}$ and CH$_{3}$OH are formed from atomic C and CO, therefore the $^{12}$C/$^{13}$C ratios of these icy molecules reflect the ratios of atomic C or CO.
After 10$^{5}$ yr when CO becomes the main carbon reservoir, the carbon isotope ratio of \ce{CH3OH} ice becomes closer to the local ISM value, reflecting that of CO in the gas phase. Meanwhile, the formation of \ce{CH4} ice becomes negligible since most atomic C is converted to CO after 10$^{5}$ yr. Therefore, \ce{CH4} ice keeps the high carbon isotope ratio in the same as before 10$^{5}$ yr.
Consequently, the $^{12}$C/$^{13}$C ratios of dominant icy molecules exhibit a bimodal profile: depletion of $^{13}$C or similar to or slightly enriched in $^{13}$C (by $\sim$ 10 $\%$) compared to the local ISM value. 

COMs are formed from simple molecules. Before around 10$^{5}$ yr when C$^{+}$ and atomic C are the main carbon reservoir, COMs (e.g. CH$_{3}$CHO) and their ices are mainly formed from $^{13}$C-poor molecules, while after around 10$^{5}$ yr when CO is main carbon reservoir, COMs (e.g. CH$_3$OCH$_3$) and their ice are formed from slightly $^{13}$C-rich CO or CH$_3$OH on the grain surface or adsorbed on grains after the formation in the gas phase.
The $^{12}$C/$^{13}$C ratios of COMs follow those of simple molecules and exhibit a bimodal profile.  
Some molecules including COMs (e.g. \ce{CH3OCHO} and HCOOH) are formed from both $^{13}$C-poor species and $^{13}$C-rich species. For example, \ce{CH3OCHO} is formed from \ce{H2CO}. \ce{H2CO} is formed from CO ice on the grain surface and \ce{CH3} in the gas phase, so the $^{12}$C/$^{13}$C ratios of these molecules including COMs show the intermediate value between the bimodal profile.

\subsubsection{The Collapse Phase}
In the collapse phase (see Fig.\ref{fig: physicalmodel}), icy molecules on the grain surface sublimate into the gas phase at their sublimation temperatures, which depend on their binding energies, and the rest in the bulk ice mantle are trapped in water ice and sublimate at $\sim$ 120 K together with water ice. 
Moreover, some species are additionally produced by grain surface reactions.
Figure \ref{fig: collapse_chemistry} shows the temporal variation of abundances and $^{12}$C/$^{13}$C ratios of selected species in the collapse phase. 
For CH$_4$, the binding energy is set to be 1300 K, CH$_4$ ice on the surface sublimates around 20 K (t$\mathrm{_{final}}$ - t$\mathrm{_{core}}$ $\sim$ 2 $\times$ 10$^{4}$ yr) and the gas phase CH$_4$ abundance increases to almost 1 $\times$ 10$^{-6}$. After that, whole CH$_4$ ice sublimates at $\sim$ 120 K (t$\mathrm{_{final}}$ - t$\mathrm{_{core}}$ $\sim$ 10$^{2}$ yr) together with water ice.
The $^{12}$C/$^{13}$C ratios of abundant molecules (molecular abundance of $\sim$ 10$^{-5}$) molecules such as CH$_4$ and CH$_3$OH after water ice sublimation (t$\mathrm{_{final}}$ - t$\mathrm{_{core}}$ $\sim$ 10$^{2}$ yr) reflect those of ice formed during the static phase. 
After water ice sublimates at $\sim$ 120 K, the $^{12}$C/$^{13}$C ratio of sublimated CH$_4$ becomes significantly depleted in $^{13}$C as well as that of CH$_4$ ice while that of sublimated \ce{CH3OH} becomes enriched in $^{13}$C as well as that of \ce{CH3OH} ice.

On the other hand, some icy COMs (e.g. CH$_3$CHO and \ce{CH3OCHO}) are produced via radical-radical reactions on the grain surface during the collapse phase.
The abundances of the additionally formed ices are equivalent to or exceeding the abundances of the icy COMs formed during the static phase.
Therefore, the $^{12}$C/$^{13}$C ratios of some sublimated COMs (e.g. CH$_3$CHO and \ce{CH3OCHO}) are different from those of their ice formed in the static phase. The $^{12}$C/$^{13}$C ratios of icy molecules formed during the collapse phase depend on the $^{12}$C/$^{13}$C ratios of reactants such as radicals. For example, the \ce{CH3OCHO} ice is formed from $^{13}$C-enriched HCO radical ($^{12}$C/$^{13}$C $\sim$ 30) during the collapse phase, so the $^{12}$C/$^{13}$C ratio of \ce{CH3OCHO} ice decreases, and eventually the $^{12}$C/$^{13}$C ratio of sublimated \ce{CH3OCHO} is different from that of ice formed in the static phase.
Consequently, the $^{12}$C/$^{13}$C ratios of some COMs after water ice sublimation are essentially a mixture of those formed during the static phase and the collapse phase, sometimes closer to the $^{12}$C/$^{13}$C ratios of icy COMs formed during the collapse phase rather than those in the static phase. Moreover, the formation of a part of COMs proceeds even in the gas phase via ion-neutral reactions after the sublimation, resulting in distinct carbon isotope ratios compared to those in ice before sublimation. The detailed explanation for other complex molecules is presented in Section \ref{subsec: species}.

\begin{figure}[t!]
\centering
\includegraphics[height=1.2\hsize]{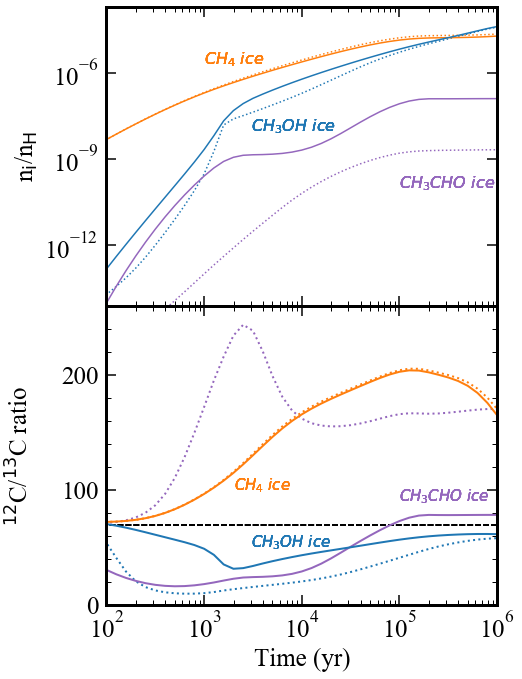}
\caption{Temporal variation of the molecular abundances and $^{12}$C/$^{13}$C ratios of CH$_4$ ice, CH$_3$OH ice, and CH$_3$CHO ice during the static phase for the base model (dotted lines) and the model with the direct C-atom addition reactions (solid lines). 
\label{fig: AbchangebyER}}
\end{figure}

\begin{figure}[t]
\centering
\includegraphics[height=.9\hsize]{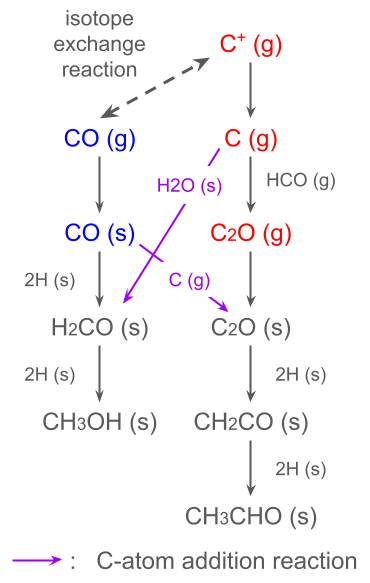}
\caption{Schematic formation pathways of \ce{CH3OH} and \ce{CH3CHO} with the direct C-atom addition reaction. Gas phase species (g) link to the icy species on the grain surface (s). The dashed arrow indicates the isotope exchange reaction. Red-colored species are depleted in $^{13}$C, while blue-colored species are enriched in $^{13}$C. Purple arrows indicate the direct C-atom addition reaction. Species 2H indicate consecutive hydrogenation reactions on the grain surfaces.
\label{fig: er}}
\end{figure}

\begin{figure}[ht!]
\centering
\includegraphics[height=1.2\hsize]{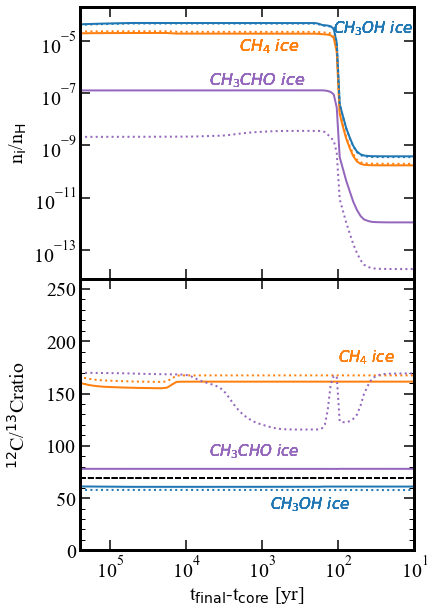}
\caption{Same as Figure \ref{fig: AbchangebyER} but during the collapse phase.
\label{fig: afterER}}
\end{figure}

\subsection{The effect of the direct C-atom addition reactions} \label{sec: insertion}

Based on recent laboratory experiments and quantum chemical calculations, we incorporate the two direct C-atom addition reactions by the ER mechanism; Eq.(\ref{eq:2}) \citep{2021Molp,2021Potapov} and Eq.(\ref{eq:3}) \citep{2022Fedoseev}.
Figure \ref{fig: AbchangebyER} shows the temporal variation of abundances and $^{12}$C/$^{13}$C ratios of some icy species in the static phase without and with the direct C-atom addition reactions.
In the model with the direct C-atom addition reactions, the abundances of icy CH$_{3}$OH and CH$_3$CHO increase compared to the base model.
On the other hand, the abundance of CH$_4$ ice, which is mainly formed via hydrogenation reactions of atomic C on the grain surfaces, slightly decreases compared to the base model (by $\sim$ 15 $\%$).
The $^{12}$C/$^{13}$C ratios of icy CH$_3$OH and CH$_{3}$CHO in the model with the direct C-atom addition reactions are closer to the ISM value compared to those in the base model. 
CH$_{3}$OH ice is formed via hydrogenation reactions of $^{13}$C-enriched CO ice on the grain surfaces in the base model. In contrast, in the direct C-atom addition model CH$_{3}$OH ice is additionally formed from slightly $^{13}$C-depleted atomic C via the reaction in Eq.(\ref{eq:2}) and the subsequent hydrogenation reactions on the grain surfaces. CH$_{3}$CHO ice is formed from $^{13}$C-depleted C$^{+}$ and atomic C via gas-phase reaction with HCO and the subsequent hydrogenation reactions on grain surfaces in the base model. In contrast, in the model with the direct C-atom addition reactions CH$_{3}$CHO ice is additionally formed from slightly $^{13}$C-depleted atomic C and $^{13}$C-enriched CO via the reaction in Eq.(\ref{eq:3}) followed by a sequence of hydrogenation reactions on the grain surfaces. These formation processes make differences in carbon isotope ratios of CH$_{3}$OH and CH$_{3}$CHO between the models with and without the direct C-atom addition reactions.

Figure \ref{fig: er} shows formation pathways of CH$_3$OH and CH$_{3}$CHO in the models with the direct C-atom addition reactions which suggests that due to these reactions, the isotope ratios of some species could change.
Figure \ref{fig: afterER} shows the temporal variation of abundances and $^{12}$C/$^{13}$C ratios of some icy species in the collapse phase without and with the direct C-atom addition reactions.
Some icy complex molecules related to the direct C-atom addition reactions (e.g. CH$_3$CHO) have lower $^{12}$C/$^{13}$C ratios compared to those in the base model. After water ice sublimation, molecules that are formed via the direct C-atom addition reactions and following reactions, also have lower $^{12}$C/$^{13}$C ratios compared to those in the base model (see Figure \ref{fig: comparison}).
We note that \citet{2021Molp} treated water clusters as the representative ice on grain surfaces, so variations in the composition of interstellar ice could affect the branching ratio \textit{A}. 
Therefore, we additionally run models changing the branching ratio \textit{A}, and find that the results obtained with \textit{A} = 0.15 are consistent with those derived when \textit{A} is set to 0.3.
We discuss the effect of the direct C-atom addition reactions for selected individual species in Section \ref{subsec: species} and we show the results of the model with the direct C-atom addition reactions in Appendix \ref{sec: additional figures}. 

\subsection{The effect of the difference in Binding Energy between $^{12}$CO and $^{13}$CO}
If the binding energies of CO isotopologues on the dust surface are mass-dependent and thus $^{13}$CO is larger than that of $^{12}$CO, $^{12}$CO can sublimate at a lower temperature than $^{13}$CO. Consequently, CO in the gas phase would become depleted in $^{13}$C, while CO ice would become enriched in $^{13}$C.
\citet{2015Smith} theoretically investigated the mass-dependence of thermal desorption of CO and dust temperature for segregation between $^{12}$CO and $^{13}$CO in the ices. They considered the balance between the adsorption rates and desorption rates for $^{12}$C$^{16}$O and $^{13}$C$^{16}$O. As a result, assuming a 10 K difference in binding energy, the $^{12}$CO/$^{13}$CO gas ratio reaches twice as high as the elemental $^{12}$C/$^{13}$C ratio at a gas temperature of 10 K.
Moreover, \citet{2021Smith} derived binding energies of pure $^{12}$CO ice and pure $^{13}$CO ice to be 833 $\pm$ 5 K, and 846 $\pm$ 6 K respectively based on laboratory experiments. 
We additionally run models assuming that the binding energy of $^{12}$CO is 833 K and that of $^{13}$CO is 846 K for both the static phase and the collapse phase.
We find that the difference in binding energy does not affect the $^{12}$C/$^{13}$C ratios of COMs although the difference leads to the desorption rate of CO to be slightly increased, nearly 10 $\%$.

\begin{figure}[ht!]
\centering
\includegraphics[height=1.2\hsize]{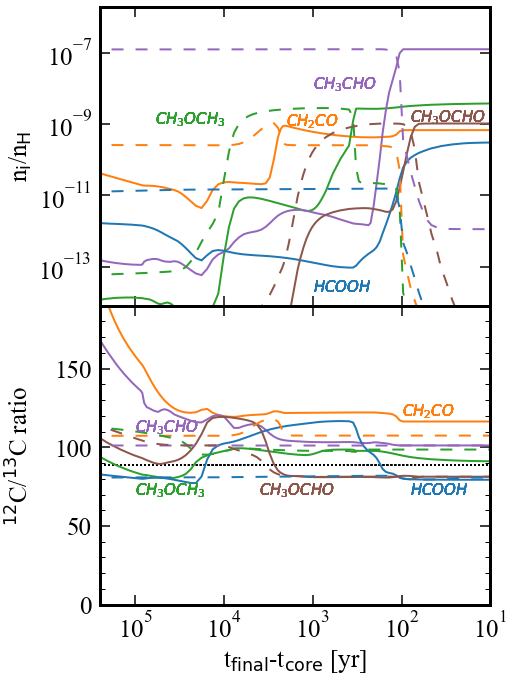}
\caption{Same as right panel of Figure \ref{fig: collapse_chemistry} but adopting the direct C-atom addition reactions and initial $^{12}$C/$^{13}$C = 89. The horizontal black dotted line in the lower panel represents the average $^{12}$C/$^{13}$C ratio of the Solar System.
\label{fig: 89}}
\end{figure}

\begin{figure}[t!]
\centering
\includegraphics[height=1.2\hsize]{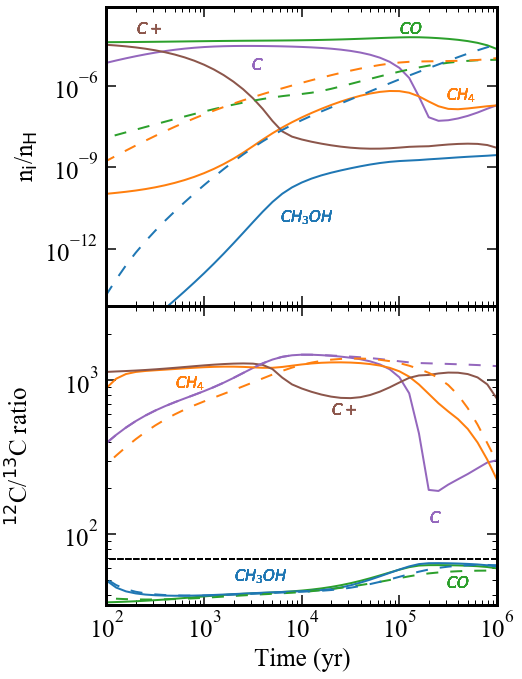}
\caption{Same as left panel of Figure \ref{fig: static_chemistry} but with the initial condition in which half of the carbon is in the form of CO, while the remaining half is in the form of \ce{C+}
\label{fig: er2_halfCO_static_chemistry}}
\end{figure}

\subsection{Dependence on the elemental $^{12}$C/$^{13}$C ratio and Initial Condition of Carbon}
In our models presented in the previous section, we assume the elemental $^{12}$C/$^{13}$C ratio is 68, which corresponds to the local ISM value \citep{2005Milam}. Here, we explore how our results depend on the assumed value of the elemental $^{12}$C/$^{13}$C ratio. Figure \ref{fig: 89} shows the temporal variation in the collapse phase with the elemental $^{12}$C/$^{13}$C ratio = 89 which is the average value in the Solar System.
We find that the $^{12}$C/$^{13}$C ratios of molecules are scaled with assumed $^{12}$C/$^{13}$C ratio.

So far we assumed that carbon is initially present as $^{12}$C$^+$ and $^{13}$C$^+$. Here, we investigate the dependence of the initial form of carbon on the carbon isotope ratios of molecules. 
We additionally run models with and without the direct C-atom addition reactions during the static phase and collapse phase where initially half of the carbon is present as $^{12}$CO and $^{13}$CO with $^{12}$CO/$^{13}$CO = 68, while the rest is present as $^{12}$C$^+$ and $^{13}$C$^+$ with $^{12}$C$^+$/$^{13}$C$^+$ = 68.  
Figure \ref{fig: er2_halfCO_static_chemistry} shows the temporal variation of abundances and $^{12}$C/$^{13}$C ratios of some selected species in the static phase without the direct C-atom addition reactions in this case.
The abundance of C$^+$ is half of that in the base model, so the abundances of molecules formed from C$^+$ such as CH$_4$ become smaller compared to the base model, and their $^{12}$C/$^{13}$C ratios become significantly elevated compared to the base model. 
For example, around 10$^{6}$ yr in the static phase, the abundance of \ce{CH4} ice is 1.0 $\times$ 10$^{-5}$, that is about 2 times smaller and the $^{12}$C/$^{13}$C ratio of \ce{CH4} is around 225, that is about 1.5 times larger compared to those in the base model, in which the abundance is 2.1 $\times$ 10$^{-5}$ and the ratio is 133.
On the other hand, CO is more abundant than in the base model, but as shown in the base model (see left panel of Fig.\ref{fig: static_chemistry}) almost carbon eventually transforms into \ce{CO} at 10$^{6}$ yr in the static phase. Therefore, the effect of initial carbon form on the molecules formed from CO is smaller than that on the molecules formed from C$^+$.
Around 10$^{6}$ yr in the static phase, the abundance of \ce{CH3OH} ice is 3.0 $\times$ 10$^{-5}$ and the $^{12}$C/$^{13}$C ratio of \ce{CH3OH} is 61, those values are similar to those in the base model, in which the abundance is 3.7 $\times$ 10$^{-5}$ and the ratio is 60.
Some complex molecules formed from both $^{13}$C-poor and $^{13}$C-rich species via radical-radical reactions or the direct C-atom addition reactions are not significantly affected by an initial form of carbon. However, complex molecules mainly formed from $^{13}$C-poor species such as \ce{C3H4} show depletion in $^{13}$C compared to the base model. 
Therefore, if we consider these molecules, we need to account for the influence of the initial form of carbon on the carbon isotope ratios.

\begin{figure*}[ht!]
\hspace*{-0.5cm}
    \begin{tabular}{cc}
      \begin{minipage}[t]{0.45\hsize}
        \centering
        \includegraphics[keepaspectratio, scale=0.5]{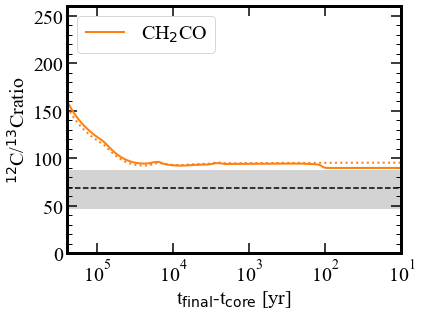}
      \end{minipage} &
      \begin{minipage}[t]{0.45\hsize}
        \centering
        \includegraphics[keepaspectratio, scale=0.5]{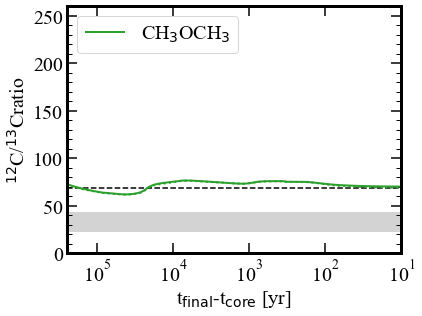}
      \end{minipage} \\
   
      \begin{minipage}[t]{0.45\hsize}
        \centering
        \includegraphics[keepaspectratio, scale=0.5]{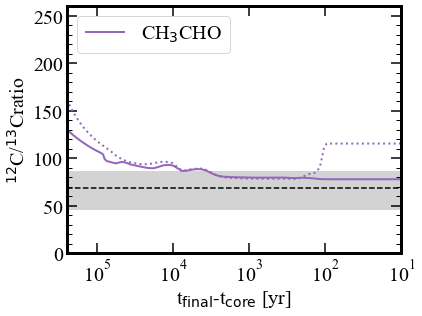}
      \end{minipage} &
      \begin{minipage}[t]{0.45\hsize}
        \centering
        \includegraphics[keepaspectratio, scale=0.5]{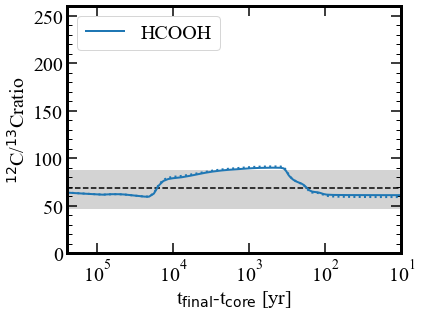}
      \end{minipage} \\

      \begin{minipage}[t]{0.45\hsize}
        \centering
        \includegraphics[keepaspectratio, scale=0.5]{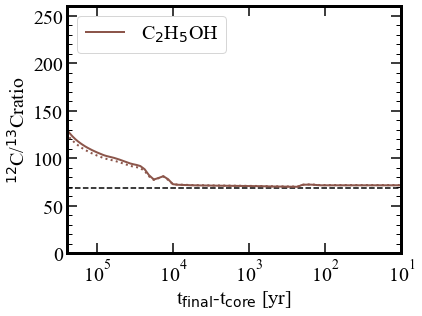}
      \end{minipage} &
      \begin{minipage}[t]{0.45\hsize}
        \centering
        \includegraphics[keepaspectratio, scale=0.5]{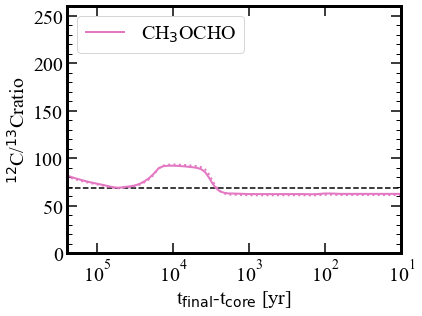}
      \end{minipage} 
      
    \end{tabular}
     \caption{Temporal variation of $^{12}$C/$^{13}$C ratios of some organic molecules for the base model (dotted lines) and the model with the direct C-atom addition reactions (solid lines). The black horizontal dashed line is for the average $^{12}$C/$^{13}$C ratio of local ISM. The observations in IRAS16293-2422B are represented in light gray shaded regions. The observations of \ce{C2H5OH} and \ce{CH3OCHO} have large uncertainties, so we refrain from comparing our results with the observations.}
     \label{fig: comparison}
  \end{figure*}

\begin{table*}[hbtp]
  \caption{$^{12}$C/$^{13}$C ratio for some organic molecules in IRAS16293-2422B and our modeling results.}
  \label{comparison}
\begin{center}
  \begin{tabular}{cccc}
    \hline \hline
    Species  & Observation$^a$  & Base Model$^b$ & C-atom Addition$^b$ \\
    \hline 
    CH$_2$CO  & 68 $\pm$ 20$^c$  & 95 & 89 \\
    CH$_3$CHO  & 67 $\pm$ 20 & 115 & 78 \\
    HCOOH  & 68 $\pm$ 20$^c$ & 59 & 61 \\
    CH$_3$OCH$_3$  &  34 $\pm$ 10 &  68 & 70 \\
    \hline
  \end{tabular}
\end{center}
    $^{(a)}$ \citet{2018Jorgensen}. The uncertainties on the ratios are 30$\%$ from error propagation of the noise in the observations. 
    $^{(b)}$ carbon isotope ratios of gas-phase molecules at 30.6 au from the core center in our models are listed.
    $^{(c)}$ The ISM value is used, which is consistent with the carbon isotope ratios derived using a few optically thin lines of main isotopologues of \ce{CH2CO} and \ce{HCOOH}.
  \label{table: comparison1}
\end{table*}

\section{Discussion} \label{sec: Discussion}

\subsection{Discussion of Individual Species} \label{subsec: species}

\subsubsection{CH$_2$CO}
In our base model, CH$_2$CO gas is depleted in $^{13}$C, and $^{12}$C/$^{13}$C $\sim$ 95 after water ice sublimation (Fig. \ref{fig: collapse_chemistry}). This value is set by the sublimation of CH$_2$CO ice. CH$_2$CO ice is mainly formed from C$_2$ ice on the grain surface via a reaction with atomic O followed by a sequence of hydrogen addition reactions on the grain surfaces in early time (t$\mathrm{_{final}}$ - t$\mathrm{_{core}}$ $\sim$ 10$^4$ yr) of the collapse phase. C$_2$ ice is mainly formed from C$_2$H$_2$ ice or via the adsorption of gaseous C$_2$, which become depleted in $^{13}$C ($^{12}$C/$^{13}$C $\sim$ 100) due to the reaction in Eq.(\ref{eq:1}), around 10$^6$ yr in the static phase.
Therefore, the $^{12}$C/$^{13}$C ratio of sublimated CH$_2$CO is depleted in $^{13}$C as well.
In the model with the direct C-atom addition reactions, the formation of CH$_2$CO ice is much more efficient in the static phase compared to the base model. CH$_2$CO ice is formed from C$_2$O ice on the grain surface via a sequence of hydrogen addition reactions. C$_2$O ice is formed by the direct C-atom addition reaction, Eq.(\ref{eq:3}). 
So, the $^{12}$C/$^{13}$C ratio of CH$_2$CO ice ($^{12}$C/$^{13}$C $\sim$ 80) is lower compared to the base model ($^{12}$C/$^{13}$C $\sim$ 180) at 10$^6$ yr in the static phase as well as \ce{CH3CHO} ice discussed in Section \ref{sec: insertion}. Additionally, $^{13}$C-depleted CH$_2$CO ice is also formed in the collapse phase as in the base model. 
At the CH$_2$CO sublimation temperature ($\sim$ 40 K, t$\mathrm{_{final}}$ - t$\mathrm{_{core}}$ $\sim$ 2 $\times$ 10$^3$ yr), the CH$_2$CO ice on the grain surface are desorbed into the gas phase and CH$_2$CO gas becomes depleted in $^{13}$C. At water sublimation temperature ($\sim$ 120 K, t$\mathrm{_{final}}$ - t$\mathrm{_{core}}$ $\sim$ 10$^2$ yr), CH$_2$CO ice in the mantle phase, which is mainly formed in the static phase and enriched in $^{13}$C, is evaporated into the gas phase and CH$_2$CO gas eventually turned into slightly enriched in $^{13}$C ($^{12}$C/$^{13}$C = 89, Fig. \ref{fig: er2_collapse_chemistry}).

\subsubsection{CH$_3$CHO}
Our base model shows that the $^{12}$C/$^{13}$C ratio of sublimated CH$_3$CHO is significantly depleted in $^{13}$C ($^{12}$C/$^{13}$C $\sim$ 115). CH$_3$CHO ice is formed from $^{13}$C-depleted C$^{+}$ and atomic C before 10$^5$ yr in the static phase as shown in Fig. \ref{fig: er}. In addition, in early time (t$\mathrm{_{final}}$ - t$\mathrm{_{core}}$ $\sim$ 10$^3$ yr, Fig. \ref{fig: static_chemistry}) of the collapse phase, CH$_3$CHO ice is formed from CH$_2$CO ice via a sequence of hydrogenation reactions or a radical-radical reaction between CH$_3$ and HCO on the grain surfaces. These reactants are slightly depleted in $^{13}$C ($^{12}$C/$^{13}$C $\sim$ 100). Therefore, sublimated CH$_3$CHO has lower $^{12}$C/$^{13}$C ratio compared to the CH$_3$CHO ice formed in the static phase (Fig. \ref{fig: collapse_chemistry}). 
In the model with the direct C-atom addition reactions, CH$_3$CHO ice is mainly formed in the static phase due to the direct C-atom addition reaction, Eq.(\ref{eq:3}), and has lower $^{12}$C/$^{13}$C ratio compared to the base model without the direct C-atom addition reactions (see Section \ref{sec: insertion}).
Unlike CH$_2$CO, CH$_3$CHO ice mainly formed in the static phase, so the $^{12}$C/$^{13}$C ratio of sublimated CH$_3$CHO is equal to that of CH$_3$CHO ice which is formed in the static phase and has lower $^{12}$C/$^{13}$C ratio compared to the base model without the direct C-atom addition reactions. Therefore, the direct C-atom addition reactions decrease the $^{12}$C/$^{13}$C ratio of CH$_3$CHO to $\sim$ 80.

In this work, we introduced two hydrogenation reactions on grain surfaces; CH$_2$CO ice + H ice $\to$ CH$_3$CO ice and CH$_3$CO ice + H ice $\to$ CH$_3$CHO ice, following \citet{2015Ruaud}. The hydrogen addition to CH$_2$CO ice is the dominant pathway for CH$_3$CHO ice formation. This pathway also plays an important role in keeping low carbon isotope ratio of CH$_3$CHO in the direct C-atom addition model since C$_2$O is formed from $^{13}$C-enriched CO and then the subsequent hydrogenation reactions of C$_2$O on the grain surfaces forms CH$_3$CHO. Without this hydrogenation pathway of CH$_2$CO, the $^{12}$C/$^{13}$C ratio of CH$_3$CHO gas is significantly depleted in $^{13}$C ($^{12}$C/$^{13}$C $\sim$ 120) after water is sublimated even though we incorporate the direct C-atom addition reactions.

\subsubsection{CH$_3$OCH$_3$}
In our base model, CH$_3$OCH$_3$ ice is mainly formed in the collapse phase from $^{13}$C-poor CH$_3$ and $^{13}$C-rich CH$_3$O radical, so the $^{12}$C/$^{13}$C ratio of CH$_3$OCH$_3$ ice is intermediate value, $\sim$ 75 before the sublimation. In addition, CH$_3$OCH$_3$ is formed in the hot gas phase via proton transfer. The gaseous CH$_3$OCH$_3$ is formed from the reaction between CH$_3$OH and CH$_3$OH$_2$$^+$ via proton transfer. The $^{12}$C/$^{13}$C ratio gradually gets close to that of CH$_3$OH after the sublimation, and decreases to $\sim$ 68. 
In the model with the direct C-atom addition reactions, the $^{12}$C/$^{13}$C ratio of CH$_3$OCH$_3$ is similar to that in the base model because CH$_3$OCH$_3$ is mostly formed during the collapse phase and the $^{12}$C/$^{13}$C ratio changes with time.
Both in the models with and without the direct C-atom addition reactions, the $^{12}$C/$^{13}$C ratio of CH$_3$OCH$_3$ becomes closer to the ISM value after water ice is sublimated.

\subsubsection{HCOOH} 
In our base model, HCOOH ice is formed from CO in the static phase, so the ice is slightly enriched in $^{13}$C (by $\sim$ 10 $\%$) at t$\mathrm{_{core}}$ $\sim$ t$\mathrm{_{final}}$. The $^{12}$C/$^{13}$C ratio of sublimated HCOOH is also slightly enriched in $^{13}$C. After the sublimation gaseous HCOOH is additionally formed from H$_2$CO, which originates from CO, via OH + H$_2$CO $\rightarrow$ HCOOH. Thus, the $^{12}$C/$^{13}$C ratio of HCOOH doesn't change and remains close to that of CO.
In the model with the direct C-atom addition reactions, the $^{12}$C/$^{13}$C ratio of HCOOH is similar to that in the base model since HCOOH ice is mainly formed from CO during the later static phase ($\sim$ 10$^{5}$ yr).

\subsubsection{\ce{C2H5OH}}
In our model, the $^{12}$C/$^{13}$C ratio of \ce{C2H5OH} gas is $\sim$ 69 after water ice sublimation. This value is set by the sublimation of \ce{C2H5OH} ice which is mainly formed via radical-radical reaction on the grain surface during the collapse phase.
The formation of \ce{C2H5OH} ice involves the rection between $^{13}$C-poor \ce{CH3} ice and slightly $^{13}$C-rich \ce{CH2OH} ice. As a result, the $^{12}$C/$^{13}$C ratio of \ce{C2H5OH} aligns with the local ISM value. This remains the case even after including the direct C-atom addition reactions.

\subsubsection{\ce{CH3OCHO}}
In our model, the $^{12}$C/$^{13}$C ratio of \ce{CH3OCHO} gas is $\sim$ 60 after water ice sublimation. This value is set by the sublimation of \ce{CH3OCHO} ice which is mainly formed via radical-radical reaction on the grain surface during the collapse phase. 
\ce{CH3OCHO} ice is formed via radical-radical reaction between $^{13}$C-poor \ce{CH3O} ice and $^{13}$C-rich HCO ice, so the $^{12}$C/$^{13}$C ratio of \ce{CH3OCHO} is the intermediate value ($\sim$ 60) regardless of whether the direct C-atom addition reactions are included.

\subsection{Comparisons with Observations of IRAS16293-2422B}

IRAS 16293-2422B is a low-mass protostar harboring a hot corino. Icy grains accrete towards the central hot region surrounding the protostar, and the bulk water ice sublimates at 100 - 200 K. As a result, the composition of the hot corino region is thought to be determined by the ice sublimation. 
In our model, the region where water ice sublimates, located inside approximately 100 au from the core center, is larger than the angular resolution of the PILS survey, which is a 60 au diameter. Therefore, this region is distinctly resolved.  
We compare our modeling results to the observational data for the IRAS 16293-2422B, which are obtained by the ALMA-PILS line surveys \citep{2016Jorgensen,2018Jorgensen}. 

In Figure \ref{fig: comparison} and Table \ref{table: comparison1}, our results are compared with the observations. 
We note that the physical model adopted in our work reproduces moderately well the observations of IRAS 16293-2422 (e.g., Figure 1 of \citealp{2014Wakelam}).
Table \ref{table: comparison1} compares the observations and the results of our calculations at t$\mathrm{_{core}}$ $\sim$ t$\mathrm{_{final}}$ for the base model and the model with the direct C-atom addition reactions. In the base model, sublimated CH$_2$CO, CH$_3$CHO, and CH$_3$OCH$_3$ are more depleted in $^{13}$C than the observations while HCOOH is similar to the observations. In the model with the direct C-atom addition reactions, CH$_2$CO and CH$_3$CHO have lower $^{12}$C/$^{13}$C ratios compared to those in the base model and are similar to the observational data.
This suggests that the direct C-atom addition reactions could play a significant role in the formation of observed organic molecules.
However, regardless of the direct C-atom addition reactions, CH$_3$OCH$_3$ remains close to the ISM value and the calculated carbon isotope ratio is larger than the observation ($\sim$ 34).
We need more investigation to reproduce the observed carbon isotope ratio of CH$_3$OCH$_3$.

As noted in \citet{2018Jorgensen} many of the observed lines of main isotopologues of CH$_2$CO and HCOOH are optically thick, so the column densities of these main isotopologues were derived from the optically thin $^{13}$C isotopologues lines assuming a standard $^{12}$C/$^{13}$C ratio (= 68). 
The derived column densities of \ce{CH2CO} and HCOOH are consistent with the observations of a few optically thin transition lines of main isotopologues.

\begin{figure}[ht!]
\plotone{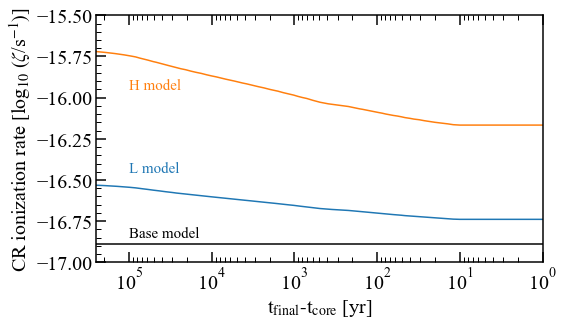}
\caption{Temporal variation of the CR ionization rate per $\mathrm{H_2}$ for the base model (black line), H model (orange line), and L model (blue line) during the collapse phase.
\label{fig: crprofile}}
\end{figure}

\begin{figure*}[ht!]
\hspace*{-1.5cm} 
    \begin{tabular}{ccc}
      \begin{minipage}[t]{0.3\hsize}
        \centering
        \includegraphics[keepaspectratio, scale=0.31]{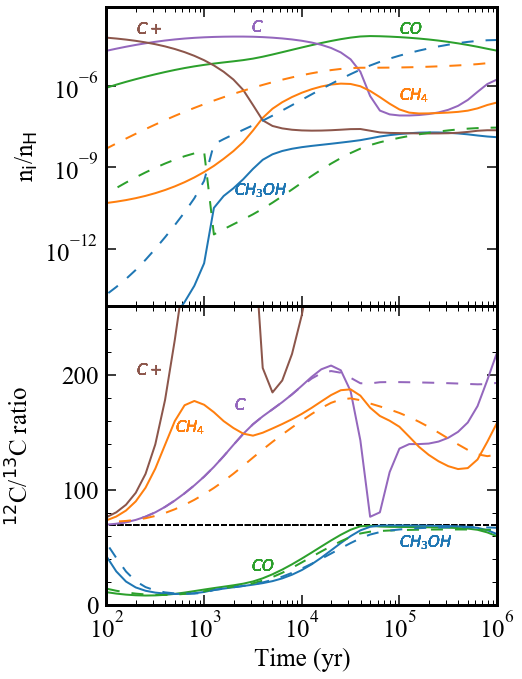}
      \end{minipage} &
      \begin{minipage}[t]{0.3\hsize}
        \centering
        \includegraphics[keepaspectratio, scale=0.31]{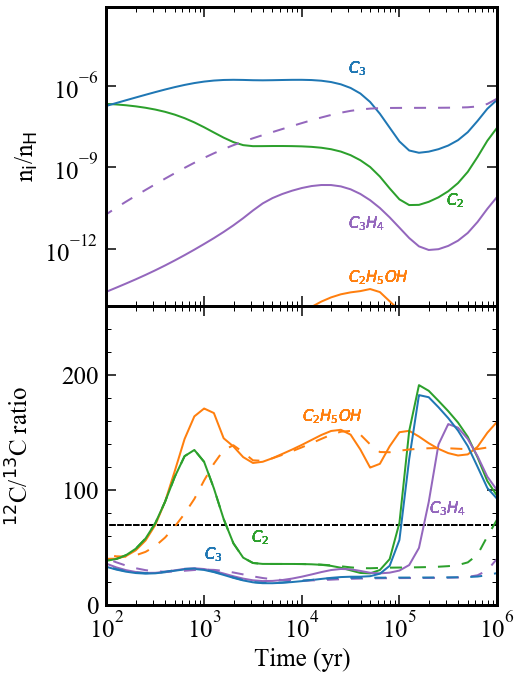}
      \end{minipage} &
      \begin{minipage}[t]{0.3\hsize}
        \centering
        \includegraphics[keepaspectratio, scale=0.31]{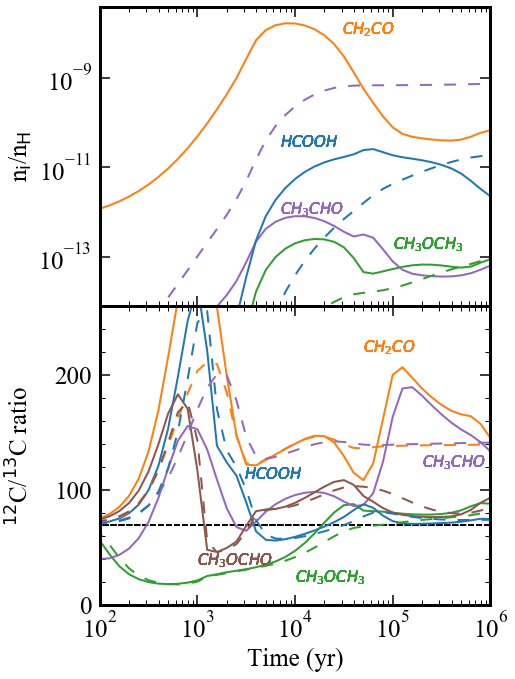}
      \end{minipage} 
    \end{tabular}
     \caption{Same as Figure \ref{fig: static_chemistry} but for the H model for CR ionization rate}
     \label{fig: static_Hchemistry}
  \end{figure*}

\begin{figure*}[ht!]
\hspace*{-1.4cm} 
    \begin{tabular}{ccc}
      \begin{minipage}[t]{0.3\hsize}
        \centering
        \includegraphics[keepaspectratio, scale=0.31]{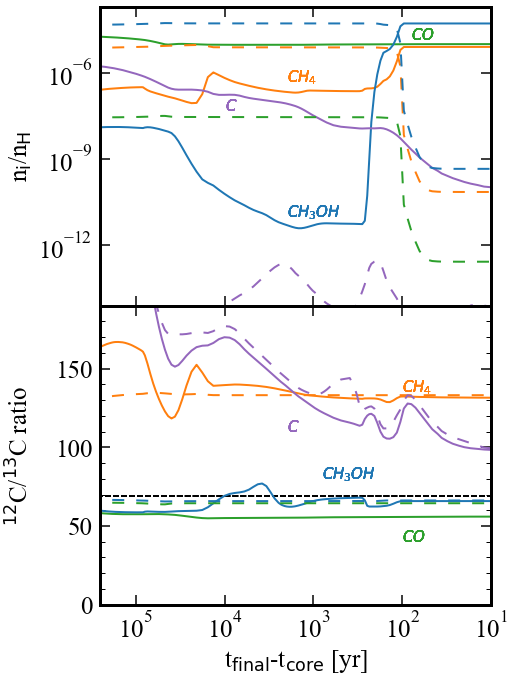}
      \end{minipage} &
      \begin{minipage}[t]{0.3\hsize}
        \centering
        \includegraphics[keepaspectratio, scale=0.31]{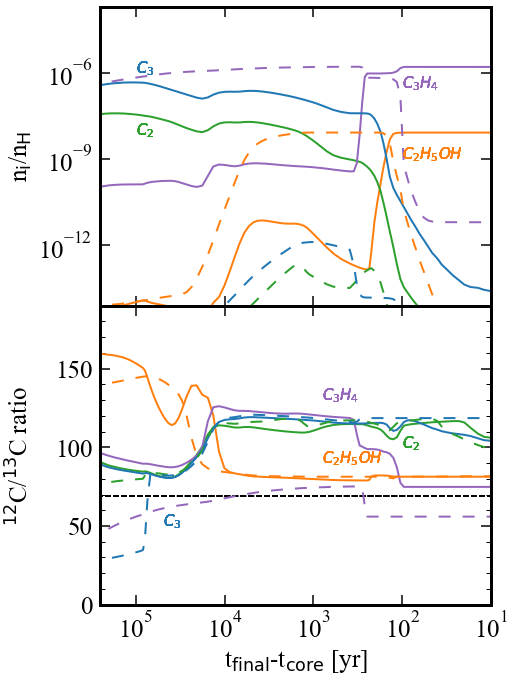}
      \end{minipage} &
      \begin{minipage}[t]{0.3\hsize}
        \centering
        \includegraphics[keepaspectratio, scale=0.31]{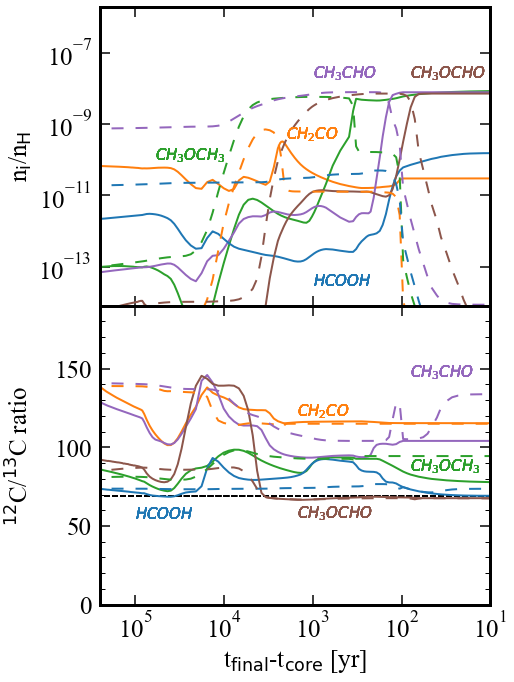}
      \end{minipage} 
    \end{tabular}
     \caption{Same as Figure \ref{fig: collapse_chemistry} but for the H model for CR ionization rate}
     \label{fig: collapse_Hchemistry}
  \end{figure*}

\subsection{The Effect of Cosmic Ray}
Since the calculated carbon isotope ratio of \ce{CH3OCH3} gas does not reproduce the observed value, we investigate the effect of cosmic ray on the $^{12}$C/$^{13}$C ratio of \ce{CH3OCH3} and other molecules.
\ce{CH3OCH3} gas is formed from ion-molecule reaction via proton transfer induced by cosmic-ray ionization after water ice sublimation in the collapse phase. Therefore, the $^{12}$C/$^{13}$C ratio of \ce{CH3OCH3} can be affected by the cosmic ray ionization rate.
In the models presented in previous sections, we assume that the cosmic-ray ionization rate of H$_2$ is constant and is 1.3$\times$10$^{-17}$ s$^{-1}$. 
Here we additionally run the astrochemical models with 
various CR ionization rates.

\subsubsection{Interstellar Cosmic Ray Flux} \label{sec: padovanicr}
We adopt the CR ionization rate model as a function of gas column density in \citet{2009P,2018P}, in which the detailed processes of energy loss and propagation of CR are taken into account.
\citet{2018P} considered two models; the L (Low spectrum) model and the H (High spectrum) model for the interstellar CRs proton spectrum. L model comes from the recent data from the extrapolation of the Voyager missions \citep{2016ApJCummings}, while the H model is from a measurement of H$_3$$^+$ in the diffuse medium \citep{2012ApJIndriolo}. 
Figure \ref{fig: crprofile} shows the temporal variation of the CR ionization rate for the model with the constant CR ionization rate of 1.3 $\times$ 10$^{-17}$ s$^{-1}$, the H model, and L model.
We run the chemical reaction network calculations using the H model and L model for CR ionization rate with and without the direct C-atom addition reactions.

Figure \ref{fig: static_Hchemistry} shows the temporal variation of molecular abundances and $^{12}$C/$^{13}$C ratios in the static phase in the H model. The higher CR ionization rate leads to the shorter timescale of ion-neutral chemical reactions including isotope exchange reactions. Gaseous CO already becomes a dominant carbon reservoir around 10$^4$ yr. Thus, the degree of the isotope exchange reactions decreases, and then the $^{12}$C/$^{13}$C ratio of CO gets closer to the local ISM value. 
The molecules formed from CO, such as CH$_3$OH, HCOOH, and CH$_3$OCH$_3$ and ices thereof mainly after 10$^5$ yr, have the $^{12}$C/$^{13}$C ratio closer to the local ISM value compared to those in the base model.
The $^{12}$C/$^{13}$C ratios of other species (e.g. CH$_3$CHO) and their ices formed from $^{13}$C-depleted C$^+$ or atomic C are smaller compared to those in the base model.
Around 10$^6$ yr higher CR ionization rate leads to atomic C being depleted in $^{13}$C because ionized and atomic C is produced by secondary photons, coming from CR-induced \ce{H2} electronic excitation and the efficiency of the isotope exchange reactions increases \citep{colzi2020}. Therefore, the molecules formed from C$^+$ or atomic C, such as gaseous CH$_4$ also become more depleted in $^{13}$C than in the base model at 10$^6$ yr. 

Figure \ref{fig: collapse_Hchemistry} shows the temporal variation of molecular abundances and $^{12}$C/$^{13}$C ratios in the collapse phase with the H model. In the early time (t$\mathrm{_{final}}$ - t$\mathrm{_{core}}$ $\sim$ 10$^4$ yr) some icy COMs (e.g. CH$_3$OCH$_3$) are produced more efficiently by the higher CR ionization rates, and these are depleted in $^{13}$C. The CR-induced UV photons dissociate stable molecules producing radicals on grain surfaces, and then stimulate the formation of COMs on the grain surface. For example, the \ce{CH3} ice on the grain surface is also depleted in \ce{^13C} since the gaseous \ce{CH4} is depleted in \ce{^13C} at the end of the static phase.
As a result, COMs formed from \ce{CH3} ice by radical-radical reactions on warm grains become more depleted in \ce{^13C} after water ice sublimation than in the base model.
Moreover, for a higher CR ionization rate, the $^{12}$C/$^{13}$C ratio of CH$_3$OCH$_3$ gas approaches that of CH$_3$OH more quickly, because the CR ionization rate promotes the formation of CH$_3$OCH$_4$$^+$ via ion-molecule reactions and then the formation of \ce{CH3OCH3} by the following dissociative recombination, CH$_3$OCH$_4$$^+$ + e$^-$ $\rightarrow$ CH$_3$OCH$_3$ form CH$_3$OCH$_3$ in the warm gas phase.
However, the $^{12}$C/$^{13}$C ratio of CH$_3$OCH$_3$ at t$\mathrm{_{core}}$ $\sim$ t$\mathrm{_{final}}$ is $\sim$ 80, which is higher than that in the base model (see Figure. \ref{fig: static_chemistry}) because the $^{12}$C/$^{13}$C ratio of CH$_3$OCH$_3$ after water ice sublimation is depleted in $^{13}$C ($^{12}$C/$^{13}$C $\sim$ 95).

\begin{figure}[ht!]
\centering
\includegraphics[height=1.2\hsize]{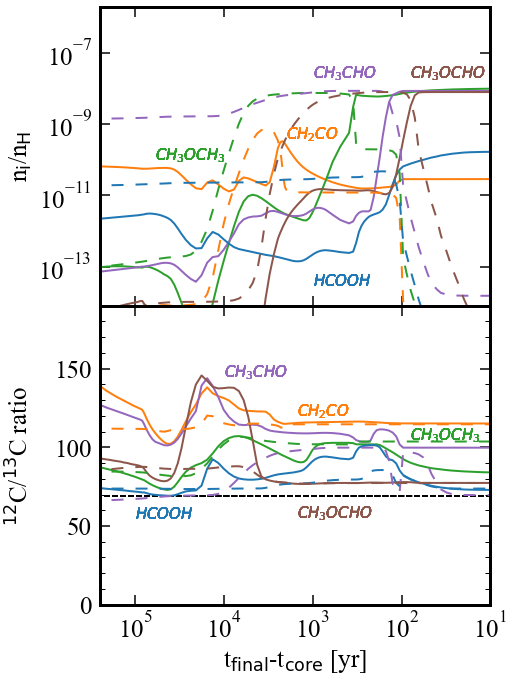}
\caption{Same as right panel of Figure \ref{fig: collapse_chemistry} but adopting the H model for CR ionization rate and the direct C-atom addition reactions.
\label{fig: chemistry_30.6Her2}}
\end{figure}

\begin{figure}[hbtp]
\centering
\includegraphics[height=1.2\hsize]{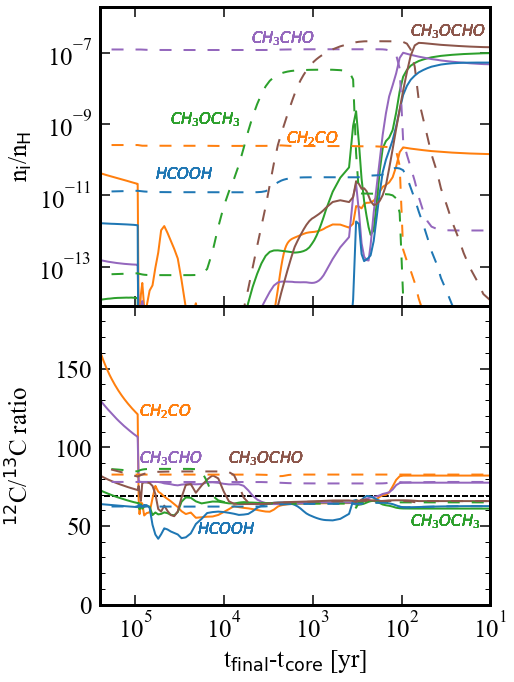}
\caption{Same as right panel of Figure \ref{fig: collapse_chemistry} but with the direct C-atom addition reactions, and 1.3$\times$10$^{-14}$ s$^{-1}$ as CR ionization rate after the protostar birth.
\label{fig: chemistry_30.6er2_Habst_birth1000cr}}
\end{figure}

We incorporate the direct C-atom addition reactions together with the H model.
Figure \ref{fig: chemistry_30.6Her2} shows the temporal variation of molecular abundances and $^{12}$C/$^{13}$C ratios in the collapse phase with the H model and the direct C-atom addition reactions. In the H model, the suppression of $^{12}$C/$^{13}$C ratios of COMs (e.g. \ce{CH3CHO}) does not occur even when considering the direct C-atom addition reactions.
Atomic C becomes CO and is depleted in the gas phase at $\sim$ 10$^4$ yr in the static phase, so the degree of the direct C-atom addition reactions become smaller, and the abundances of icy COMs produced via the direct C-atom addition reactions in the static phase becomes smaller compared to the model with the constant CR ionization rate of 1.3 $\times$ 10$^{-17}$ s$^{-1}$ and the direct C-atom addition reactions. These icy COMs are much more formed via radical-radical reactions in the collapse phase in the same as the model without the direct C-atom addition reactions. As a result, the $^{12}$C/$^{13}$C ratios of COMs in the warm gas trace those formed on warm dust grains during the collapse phase.
For example, the abundance of \ce{CH3CHO} ice at 10$^6$ yr in the static phase decreases by a factor of 100 (the molecular abundance is $\sim$ 10$^{-9}$)  (see also Figure \ref{fig: AbchangebyER}). On the other hand, the abundance of \ce{CH3CHO} ice formed from \ce{^13C}-depleted \ce{CH3} during the collapse phase is ten times larger than that formed in the static phase. As a result, the $^{12}$C/$^{13}$C ratio of \ce{CH3CHO} gas after water ice sublimation is depleted in $^{13}$C despite incorporating the direct C-atom addition reactions.
Therefore, the high CR ionization rate during the static phase is not suitable for the formation of COMs (such as \ce{CH3CHO}) in the viewpoint of their carbon isotope ratios.
For the L model, the result is similar to that of the model with the constant CR ionization rate of 1.3 $\times$ 10$^{-17}$ s$^{-1}$.

\subsubsection{Cosmic Ray Acceleration after protostar formation}
Following the results in Sect. \ref{sec: padovanicr}, in this section, we consider the effect of possible variations of CR ionization rate during the star formation.
CRs are suggested to be accelerated by the local shocks around the protostar \citep{2015Padovani,2016Padovani} such as strongly magnetized shock along the outflow or by the accretion shocks. \citet{2023Cabedo} measured abundances of molecular ions in a solar-type protostellar object and suggest high CR ionization rates of 10$^{-16}$ - 10$^{-14}$ s$^{-1}$, possibly locally accelerated at shocks. 
We run the additional model in which we set the CR ionization rate to be 1.3$\times$10$^{-14}$ s$^{-1}$ after the protostellar formation. 
We note that the CR ionization rate higher than 1.3$\times$10$^{-14}$ s$^{-1}$ destroys COMs via proton transfer by \ce{H3O+} after sublimation as suggested by \citet{2004Nomura} and eventually abundance of some molecules (e.g. \ce{CH3CHO} and \ce{CH3OCH3}) decrease to around 1 $\times$ 10$^{-11}$ at t$\mathrm{_{core}}$ = t$\mathrm{_{final}}$.
Figure \ref{fig: chemistry_30.6er2_Habst_birth1000cr} shows the temporal variation of molecular abundances and $^{12}$C/$^{13}$C ratios in the collapse phase with CR ionization rate of 1.3$\times$10$^{-14}$ s$^{-1}$ and the direct C-atom addition reactions. CH$_{3}$OCH$_{3}$ is formed from CH$_{3}$OH and CH$_{3}$OH$_{2}$$^{+}$ via proton transfer after water ice sublimation and then the $^{12}$C/$^{13}$C ratio of CH$_{3}$OCH$_{3}$ approaches that of CH$_{3}$OH and becomes enriched in $^{13}$C ($\sim$ 60). 
Consequently, the $^{12}$C/$^{13}$C ratio of CH$_{3}$OCH$_{3}$ decreases if the CR ionization rates become high only in the collapse phase. However, the ratio is still higher than the observation towards IRAS16293-2422B \citep{2018Jorgensen}. 
For other selected molecules, the $^{12}$C/$^{13}$C ratios of HCOOH, CH$_{2}$CO, CH$_{3}$CHO are similar to those in the base model with the direct C-atom addition reactions (see a right panel of Fig. \ref{fig: er2_collapse_chemistry}). 

\section{Summary} \label{sec: sum}
We investigated the carbon isotope fractionation of COMs from prestellar cores to protostellar cores by combining the chemical network model and the radiation hydrodynamical simulation. The temporal variation of the molecular abundances and $^{12}$C/$^{13}$C ratios were calculated by conducting calculations of a gas-grain chemical network within a single fluid parcel. Our main findings are the following.

\begin{enumerate}
    \item In the static prestellar phase, due to the carbon isotope exchange reactions, the molecules formed from atomic C and C$^{+}$ were depleted in $^{13}$C, while the molecules formed from CO were around 10 $\%$ enriched in $^{13}$C at the end of the static phase. COMs have various $^{12}$C/$^{13}$C ratios depending on the $^{12}$C/$^{13}$C ratios of their reactants.    
    
    \item In the collapse phase, after the protostar formation where species are sublimated to the gas phase, the $^{12}$C/$^{13}$C ratios of some sublimated species (e.g. CH$_4$ and CH$_3$OH) reflected those of their icy counterparts. On the other hand, the $^{12}$C/$^{13}$C ratios of other species are affected by chemical reactions during the collapse phase. Some COMs (e.g. CH$_3$CHO) were formed on warm grain surfaces during the collapse phase as much as in the static phase. As a result, the $^{12}$C/$^{13}$C ratios of these COMs were different from those of their icy counterparts formed in the static phase. In addition, some COMs (e.g. CH$_3$OCH$_3$) were formed in the warm gas phase after water ice sublimation, which also affects their carbon isotope ratios. Eventually, in our base model, some complex molecules (e.g. CH$_3$CHO and CH$_3$OCH$_3$) become more depleted in $^{13}$C compared to the observations.
    
    \item We incorporated the direct C-atom addition reactions into our base model and investigated the effect of these reactions on the formation and the $^{12}$C/$^{13}$C ratios of COMs.
    Direct C-atom addition reactions altered the $^{12}$C/$^{13}$C ratios of molecules, reducing the diversity of the carbon isotope ratios. This is due to the additional formation of specific species (e.g., \ce{H2CO} and \ce{C2O}), which further lead to the production of COMs. Our results reproduced the observations of the $^{12}$C/$^{13}$C ratios of some complex molecules including COMs (CH$_2$CO, CH$_3$CHO, and HCOOH) in the Class 0 source, IRAS16293-2422B, which were comparable to the elemental $^{12}$C/$^{13}$C ratio in the local ISM. However, CH$_3$OCH$_3$ in our results still shows depletion of $^{13}$C compared to the observations. 
    
    \item We investigated the effect of various CR ionization rates on carbon isotope ratios of complex molecules. 
    During the static phase, high cosmic ray (CR) ionization rates are considered unfavorable for the formation of complex organic molecules (COMs). Consequently, the formation of $^{13}$C-rich COMs through direct C-atom addition reactions becomes less efficient. Instead, $^{13}$C-depleted radicals, such as CH$_3$, tend to form COMs on warm grain surfaces during the collapse phase. On the other hand, ion-molecule reactions in the warm gas phase reduced the $^{12}$C/$^{13}$C ratio of CH$_{3}$OCH$_{3}$ after water ice sublimation. Especially, when the CR ionization rates become as high as 1.3$\times$10$^{-14}$ s$^{-1}$ only after protostar formation, this ratio becomes enriched in $^{13}$C. 
    However, the $^{12}$C/$^{13}$C ratio of CH$_{3}$OCH$_{3}$ at the end of the collapse phase was still higher than the observations.

\end{enumerate}

We thank the anonymous referees for the helpful comments to improve the manuscript.
This work is supported by JSPS and MEXT Grants-in-Aid for Scientific Research, 18H05441, 19K03910, 20H00182 (H.N.).

\bibliography{main}{}
\bibliographystyle{aasjournal}

\appendix
\setcounter{table}{0}
\renewcommand{\thetable}{A\arabic{table}}
\setcounter{figure}{0}
\renewcommand{\thefigure}{A\arabic{figure}}

\section{Review of carbon isotope exchange reactions included in the model}
In table \ref{table: exchange}, we list carbon isotope exchange reactions used in our model.
\begin{table*}[hbtp]
  \caption{Review of carbon isotope exchange reactions included in the model}
  \label{exchange}
\begin{center}
  \begin{tabular}{llcc}
    \hline \hline
    Reaction  & Rate Coefficient & $\Delta$E (K) & References \\
    \hline 
    $^{13}$\ce{C+} + \ce{CO} $\rightleftharpoons$ \ce{C+} + $^{13}$\ce{CO} & $6.6\times10^{-10}\times(\frac{T}{300})^{-0.45}$  & 34.5 & (1) \\
    $^{13}$\ce{CO} + \ce{HCO+} $\rightleftharpoons$ \ce{CO} + H$^{13}$C\ce{O+}  & 2.6$\times$10$^{-10}$$\times$($\frac{T}{300})$$^{-0.33}$ & 17.4 & (1) \\
    $^{13}$\ce{C+} + \ce{C2} $\rightleftharpoons$ \ce{C+} + $^{13}$\ce{CC}  & 1.86$\times$10$^{-9}$ & 26.4 & (2) \\
    $^{13}$\ce{C+} + \ce{C3} $\rightleftharpoons$ \ce{C+} + $^{13}$\ce{CC2}  &  1.8$\times$10$^{-9}$ &  28.0 & (2) \\
    $^{13}$\ce{C+} + \ce{CN} $\rightleftharpoons$ \ce{C+} + $^{13}$\ce{CN}  &  $3.82\times10^{-9}\times(\frac{T}{300})^{-0.4}$ &  31.1 & (1) \\
    $^{13}$\ce{C+} + \ce{CS} $\rightleftharpoons$ \ce{C+} + $^{13}$\ce{CS}  &  2.0$\times$10$^{-9}$$\times(\frac{T}{300})^{-0.4}$ &  26.3 & (2) \\
    $^{13}$\ce{C} + \ce{H2CN+} $\rightleftharpoons$ \ce{C} + \ce{H2}$^{13}$\ce{CN+}  &  1.0$\times$10$^{-9}$ &  50.0 & (2) \\
    \ce{H2}$^{13}$\ce{CN+} + \ce{HCN} $\rightleftharpoons$ \ce{H2CN+} + H$^{13}$\ce{CN}  &  2.0$\times$10$^{-9}$$\times$($\frac{T}{300})$$^{-0.5}$ &  2.9 & (2) \\   
    \ce{H2CN+} + HN$^{13}$\ce{C} $\rightleftharpoons$ \ce{H2}$^{13}$\ce{CN+} + \ce{HCN}  &  1.0$\times$10$^{-9}$$\times$($\frac{T}{300})$$^{-0.5}$ &  0 & (2) \\   
    \ce{H2CN+} + HN$^{13}$\ce{C} $\rightleftharpoons$ \ce{H2CN+} + H$^{13}$\ce{CN}  &  1.0$\times$10$^{-9}$$\times$($\frac{T}{300})$$^{-0.5}$ &  0 & (2) \\ 
    $^{13}$\ce{C} + \ce{C2} $\rightleftharpoons$ \ce{C} + $^{13}$\ce{CC}  & 3.0$\times$10$^{-10}$ & 26.4 & (2) \\
    $^{13}$\ce{C} + \ce{C3} $\rightleftharpoons$ \ce{C} + $^{13}$\ce{CC2}  &  3.0$\times$10$^{-10}$ &  28.0 & (2) \\
    $^{13}$\ce{C} + \ce{CN} $\rightleftharpoons$ \ce{C} + $^{13}$\ce{CN}  &  3.0$\times$10$^{-10}$ &  31.1 & (1) \\  
    $^{13}$\ce{C} + \ce{HCN} $\rightleftharpoons$ \ce{C} + H$^{13}$\ce{CN}  &  2.0$\times$10$^{-10}$ &  48.0 & (2) \\
    $^{13}$\ce{C} + \ce{HNC} $\rightleftharpoons$ \ce{C} + HN$^{13}$\ce{C}  &  3.0$\times$10$^{-11}$ &  33.0 & (2) \\
    $^{13}$\ce{C} + \ce{CS} $\rightleftharpoons$ \ce{C} + $^{13}$\ce{CS}  &  2.0$\times$10$^{-10}$ &  26.3 & (2) \\
    $^{13}$\ce{C} + \ce{HC3N} $\rightleftharpoons$ \ce{C} + H$^{13}$\ce{CC2N}  &  6.0$\times$10$^{-11}$ &  48.3 & (2) \\
    \hline
  \end{tabular}
\end{center}
    References. (1) \citet{2015Roueff}; (2) \citet{loison2020}.
  \label{table: exchange}
\end{table*}

\newpage
\section{Additional figures} \label{sec: additional figures}
\subsection{Carbon- and Oxygen-bearing molecules}
Figure \ref{fig: er2_static_chemistry} and \ref{fig: er2_collapse_chemistry} show the temporal variation of abundances and the $^{12}$C/$^{13}$C ratios of some molecules with the direct C-atom addition reactions in the static phase and collapse phase, respectively.

\renewcommand{\thefigure}{B\arabic{figure}}
\setcounter{figure}{0} 
\begin{figure*}[hbtp]
\hspace*{-1.5cm} 
    \begin{tabular}{ccc}
      \begin{minipage}[t]{0.3\hsize}
        \centering
        \includegraphics[keepaspectratio, scale=0.31]{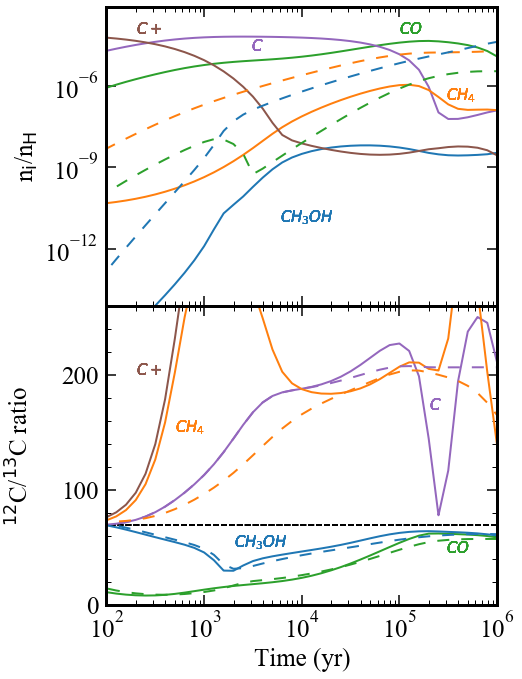}
      \end{minipage} &
      \begin{minipage}[t]{0.3\hsize}
        \centering
        \includegraphics[keepaspectratio, scale=0.31]{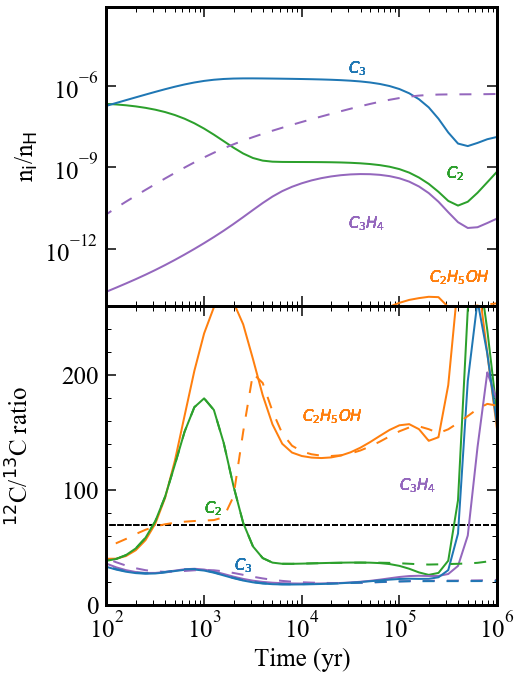}
      \end{minipage} &
      \begin{minipage}[t]{0.3\hsize}
        \centering
        \includegraphics[keepaspectratio, scale=0.31]{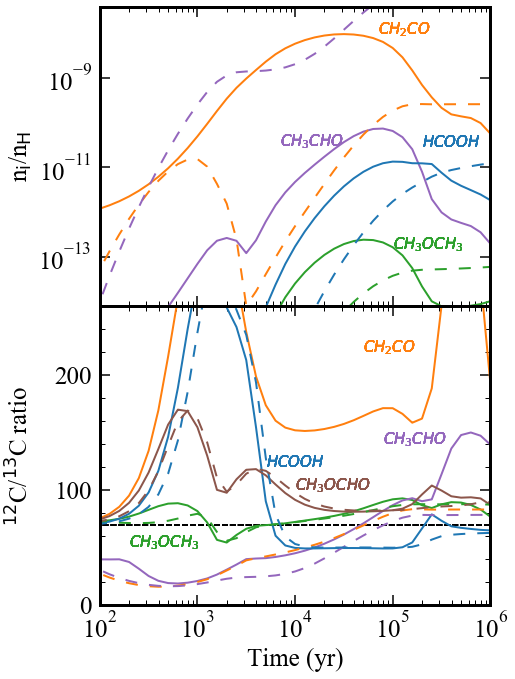}
      \end{minipage} 
    \end{tabular}
     \caption{Same as Figure \ref{fig: static_chemistry} but adopting the direct C-atom addition reactions}
     \label{fig: er2_static_chemistry}
\end{figure*}

\begin{figure*}[hbtp]
\hspace*{-1.4cm} 
    \begin{tabular}{ccc}
      \begin{minipage}[t]{0.3\hsize}
        \centering
        \includegraphics[keepaspectratio, scale=0.31]{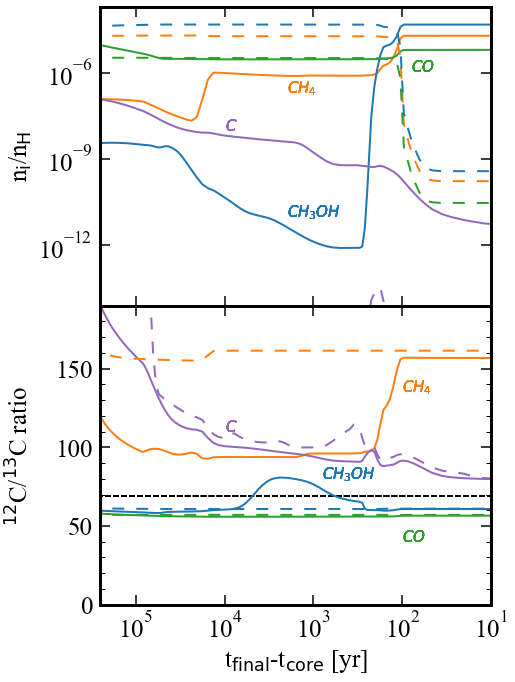}
      \end{minipage} &
      \begin{minipage}[t]{0.3\hsize}
        \centering
        \includegraphics[keepaspectratio, scale=0.31]{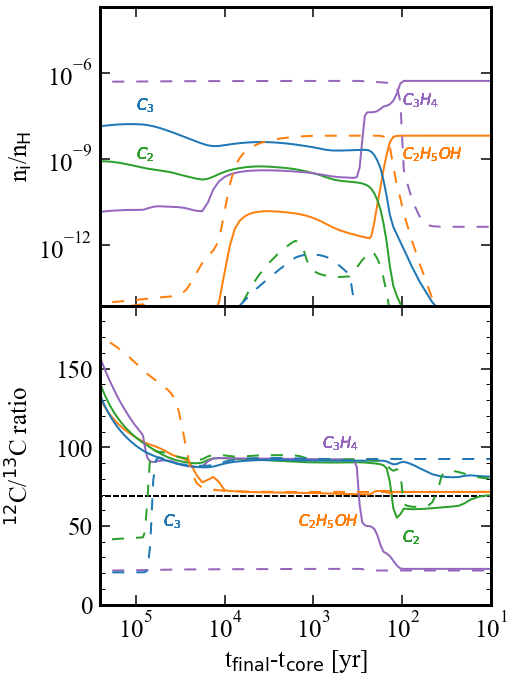}
      \end{minipage} &
      \begin{minipage}[t]{0.3\hsize}
        \centering
        \includegraphics[keepaspectratio, scale=0.31]{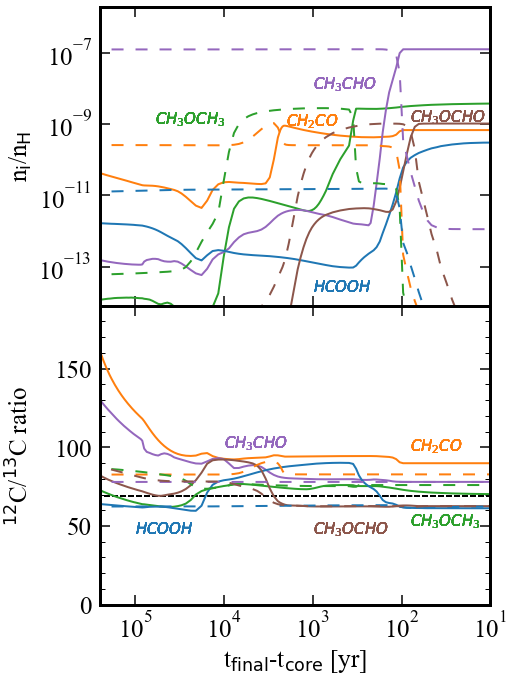}
      \end{minipage} 
    \end{tabular}
     \caption{Same as Figure \ref{fig: collapse_chemistry} but adopting the direct C-atom addition reactions.}
     \label{fig: er2_collapse_chemistry}
\end{figure*}

\begin{figure*}[hbtp]
\hspace*{-1.4cm} 
    \begin{tabular}{cc}   
      \begin{minipage}[t]{0.55\hsize}
        \centering
        \includegraphics[keepaspectratio, scale=0.4]{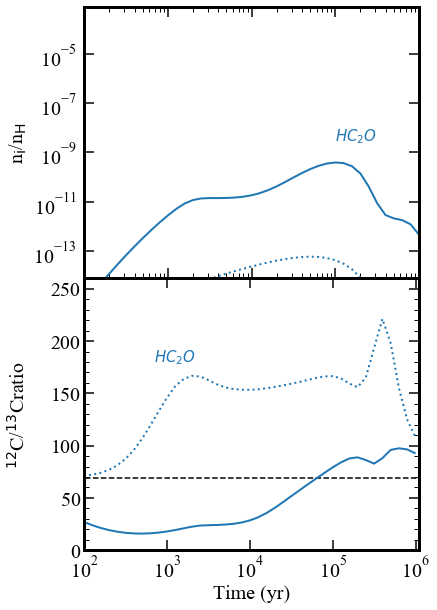}
      \end{minipage} &
      \begin{minipage}[t]{0.5\hsize}
        \centering
        \includegraphics[keepaspectratio, scale=0.4]{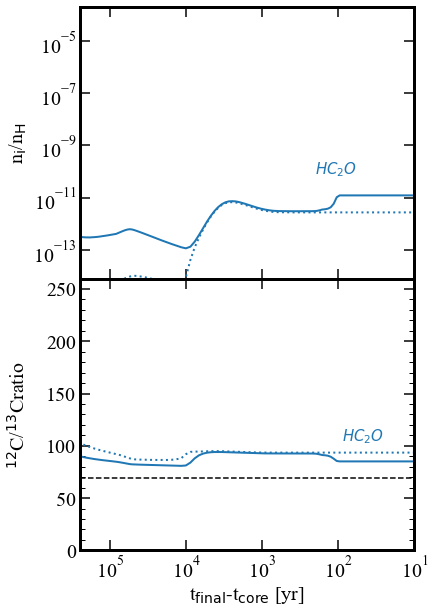}
      \end{minipage} 
    \end{tabular}
     \caption{The temporal variation of abundance and the $^{12}$C/$^{13}$C ratio of \ce{HC2O} gas for the base model (dotted lines) and the model with the direct C-atom addition reactions (solid lines). The left panel is for the static phase and the right panel is for the collapse phase.}
     \label{fig: AbchangebyERHC2O}
\end{figure*}

Figure \ref{fig: AbchangebyERHC2O} shows the temporal variation of abundance and the $^{12}$C/$^{13}$C ratio of \ce{HCCO} gas. \ce{HCCO} is formed from \ce{C2O} ice which is formed via direct C-atom addition reaction \ref{eq:3}. Therefore, after incorporating the direct C-atom addition reactions the abundance increases, and the $^{12}$C/$^{13}$C ratio decreases. 
\ce{HCCO} was observed toward the starless core \citep{2015Agundez}, where they reported the column density ratio of \ce{HCCO} and \ce{H2}, ranging from 10$^{-13}$ to 10$^{-11}$.
In our model, \ce{HCCO} gas to \ce{H_2} gas ratio at typical dense cloud age ($\sim$ 10$^5$ yr) in the static phase increases from 4.4$\times$10$^{-14}$ to 3.5$\times$10$^{-10}$ with the direct C-atom addition reaction \ref{eq:3}. This means that HCCO gas is efficiently formed via the direct C-atom addition reaction \ref{eq:3} even at low-temperature environments. The $^{12}$C/$^{13}$C ratio of HCCO gas decreases from 160 to 80 at typical dense cloud age, and from 93 to 85 after water ice sublimates. Therefore, measuring the $^{12}$C/$^{13}$C ratio of HCCO by observation could confirm whether the direct C-atom addition reactions contribute to forming COMs.

\subsection{Nitrogen-bearing molecules} \label{sec: Nspecies}
\begin{figure*}[hbtp]
\hspace*{-0.4cm} 
    \begin{tabular}{ccc}
      \begin{minipage}[t]{0.31\hsize}
        \centering
        \includegraphics[keepaspectratio, scale=0.31]{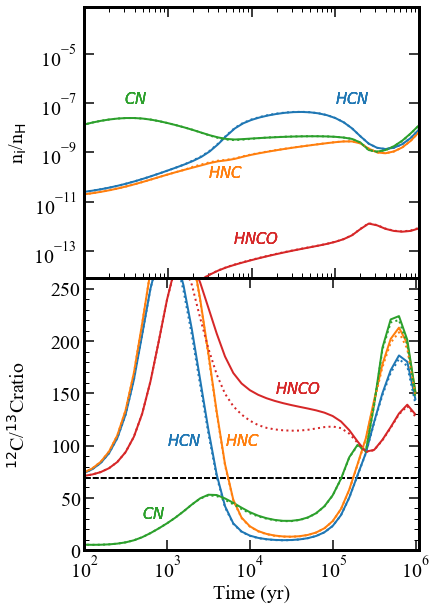}
      \end{minipage} &
      \begin{minipage}[t]{0.31\hsize}
        \centering
        \includegraphics[keepaspectratio, scale=0.31]{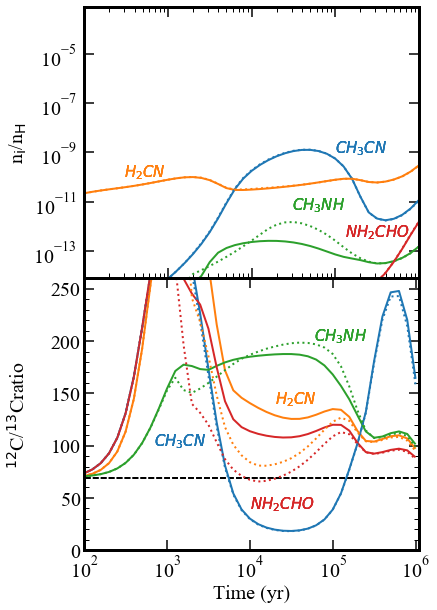}
      \end{minipage} &
      \begin{minipage}[t]{0.31\hsize}
        \centering
        \includegraphics[keepaspectratio, scale=0.31]{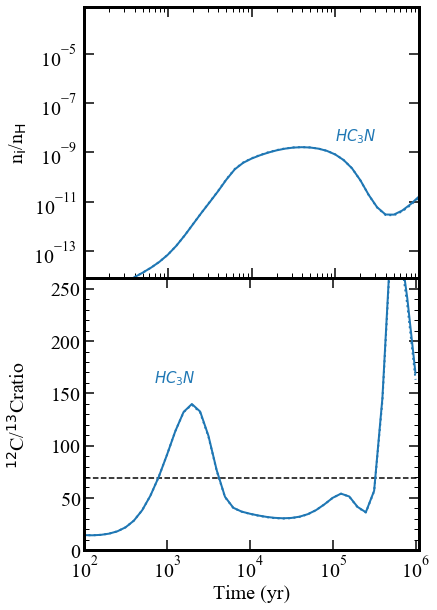}
      \end{minipage} 
    \end{tabular}
     \caption{Temporal variation of the molecular abundances and $^{12}$C/$^{13}$C ratios for gaseous species during the static phase in the model with the direct C-atom addition reactions (solid lines) and the base model (dashed lines). The horizontal black dashed line represents the average $^{12}$C/$^{13}$C ratio of local ISM.}
     \label{fig: static_others}
  \end{figure*}

\begin{figure*}[hbtp]
\hspace*{-0.4cm} 
    \begin{tabular}{ccc}
      \begin{minipage}[t]{0.31\hsize}
        \centering
        \includegraphics[keepaspectratio, scale=0.31]{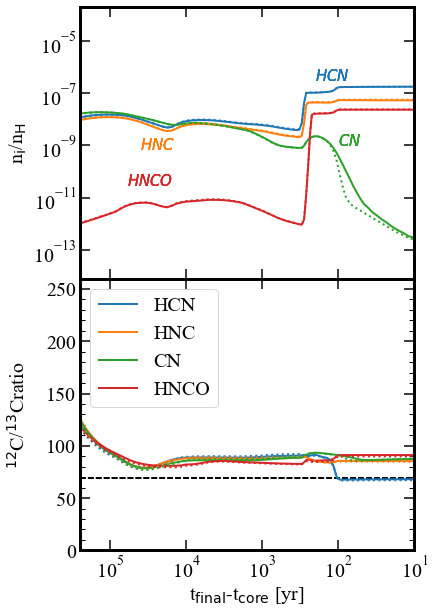}
      \end{minipage} &
      \begin{minipage}[t]{0.31\hsize}
        \centering
        \includegraphics[keepaspectratio, scale=0.31]{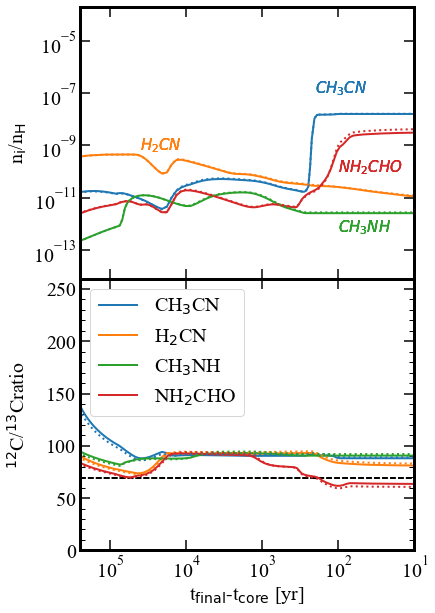}
      \end{minipage} &
      \begin{minipage}[t]{0.31\hsize}
        \centering
        \includegraphics[keepaspectratio, scale=0.31]{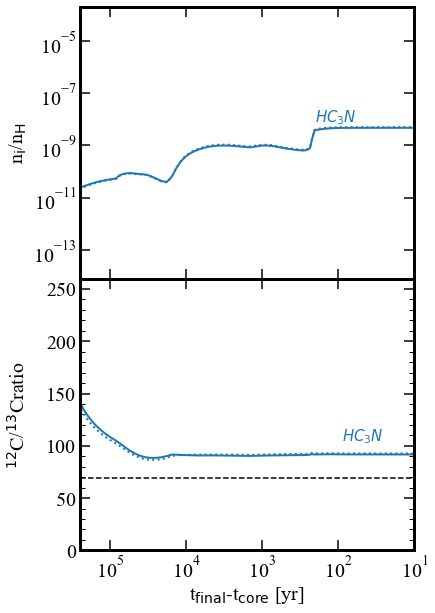}
      \end{minipage} 
    \end{tabular}
     \caption{Same as Figure \ref{fig: static_others} but for during the collapse phase.}
     \label{fig: collapse_others}
  \end{figure*}

Figures \ref{fig: static_others} and \ref{fig: collapse_others} show the temporal variation of abundances and $^{12}$C/$^{13}$C ratios of nitrogen-bearing species in the static phase and the collapse phase, respectively. 
These results in the static phase in our model are similar to the results in \citet{colzi2020} and \citet{loison2020}. The $^{12}$C/$^{13}$C ratios are affected by carbon isotope exchange reactions.
Some molecules (e.g. HNCO, \ce{H2CN}, \ce{NH2CHO} and \ce{CH3NH}) show different time variations compared to our base model at around 10$^4$ yr in the static phase due to the direct C-atom addition reaction \ref{eq:2}. These molecules are formed from \ce{H2CO} ice. In the base model \ce{HNCO}, \ce{H2CN}, and \ce{NH2CHO} are partly formed from $^{13}$C-enriched CO and \ce{H2CO}, but in the model with the direct C-atom addition reactions \ce{H2CO} is formed via Reaction \ref{eq:2} and is depleted in  $^{13}$C, similar to atomic C. Therefore, the $^{12}$C/$^{13}$C ratios of these molecules increase when the direct C-atom addition reactions are incorporated.
\ce{CH3NH} is formed from \ce{CH3} or \ce{H2CN}, so the $^{12}$C/$^{13}$C ratio changes by the direct C-atom addition reactions.
Around 10$^6$ yr, however, these differences are no longer evident because \ce{H2CO} ice is formed from CO ice via a sequence of hydrogenation reactions on the grain surface.
In the collapse phase, there is no difference between our base model and the model with the direct C-atom addition reactions.
For \ce{HC3N}, the $^{12}$C/$^{13}$C ratio depends on the position of $^{13}$C within the molecules. Therefore, we will address these aspects in future work.

\end{document}